\documentclass[12pt]{article}

\newcommand{\VersionInformation}{}  

\renewcommand{\VersionInformation}{%
  \newcount\hour \newcount\minute%
  \hour=\time \divide\hour 60%
  \minute=\hour \multiply\minute-60 \advance\minute\time\relax%
  \parbox{10cm}{\textbf{%
      Version from \today,\ \the\hour:\the\minute\\
      \texttt{\# latex \jobname.tex}
    }}%
}%

\usepackage{amsmath}
\usepackage{amsthm}
\usepackage{amsfonts}          
\usepackage{amssymb}           
\usepackage[mathscr]{eucal}
\usepackage{graphicx}
\usepackage{color}
\usepackage{hypbmsec}
\usepackage{fancyvrb}
\usepackage{sepnum}

\usepackage{rotating}
\usepackage{slashbox}
\usepackage[isu,bf]{caption}
\newlength{\xtrawidth}
\newlength{\xtraheight}
\usepackage[all]{xy}           

\usepackage[bookmarks=false]{hyperref}

\def\clap#1{\hbox to 0pt{\hss#1\hss}}

\makeatletter
  \def\adots{\mathinner{\mkern2mu\raise\p@\hbox{.}
      \mkern2mu\raise4\p@\hbox{.}\mkern1mu
      \raise7\p@\vbox{\kern7\p@\hbox{.}}\mkern1mu}}
\makeatother

\newcommand{\comma}[1]{\ensuremath{\sepnum{{.}}{{,}}{}{#1}}}

\newcommand{\Z}{\mathbb{Z}}
\newcommand{\CP}{\ensuremath{\mathop{\null {\mathbb{P}}}}\nolimits}

\newcommand{\sbar}{\ensuremath{{\bar{s}}}}

\DeclareMathOperator{\diff}{d\!}

\DeclareMathOperator{\tr}{tr}

\DeclareMathOperator{\rank}{rank}

\DeclareMathOperator{\Vol}{Vol}
\DeclareMathOperator{\dVol}{dVol}

\newcommand{\Lsheaf}{\ensuremath{\mathscr{L}}}
\newcommand{\Osheaf}{\ensuremath{\mathscr{O}}}

\newcommand{\Vsheaf}{\ensuremath{\mathscr{V}}}

\newcommand{\Kahler}{K\"ahler}

\newcommand{\Omegabar}{\overline{\Omega}}
\DeclareMathOperator{\Ric}{Ric}

{\end{list}}

{\everymath{\displaystyle\everymath{}}\array}%
{\endarray}

\newtheorem{theorem}{Theorem}

\numberwithin{equation}{section}

\newcommand{\beq}{\begin{equation}}
\newcommand{\eeq}{\end{equation}}

\newcommand{\bea}{\begin{eqnarray}}
\newcommand{\eea}{\end{eqnarray}}
\newcommand{\ba}{\begin{array}}
\newcommand{\ea}{\end{array}}
\newcommand{\eref}[1]{(\ref{#1})}
\newcommand{\rk}{\mathop{{\rm rk}}}

\newcommand{\cO}{{\cal O}}

\newcommand{\cF}{{\cal F}}

\newcommand{\cL}{{\cal L}}
\newcommand{\cV}{{\cal V}}

\begin{document}
\begin{titlepage}
  \vspace*{-2cm}
  \hfill
  \parbox[c]{5cm}{
    \begin{flushright}
    \end{flushright}
  }
  \vspace*{2cm}
  \begin{center}
    \LARGE
    Numerical Hermitian Yang-Mills Connections\\
    and Vector Bundle Stability\\
    in Heterotic Theories
  \end{center}
  \vspace*{8mm}
  \begin{center}
    \begin{minipage}{\textwidth}
      \begin{center}
        \sc 
        Lara B.\ Anderson${}^{1}$,
        Volker Braun${}^{2}$,
        Robert L.\ Karp${}^{3}$,\\ and
        Burt A.\ Ovrut${}^{1}$
      \end{center}
      \begin{center}
        \textit{
          ${}^1$Department of Physics, University of Pennsylvania\hphantom{${}^1$}\\ 
          209 South 33rd Street, Philadelphia, PA 19104-6395, U.S.A.
        }\\[1ex]
        \textit{
          ${}^2$Dublin Institute for Advanced Studies\hphantom{${}^2$}\\
          10 Burlington Road, Dublin 4, Ireland
        }\\[1ex]
        \textit{
          ${}^3$Department of Physics, Virginia Polytechnic Institute and\hphantom{${}^3$}\\
          State University, Blacksburg, VA 24061, U.S.A.
        }
      \end{center}
    \end{minipage}
  \end{center}
  \vspace*{\stretch1}
  \begin{abstract}
  A numerical algorithm is presented for explicitly computing the gauge connection on slope-stable holomorphic vector bundles on Calabi-Yau manifolds. To illustrate this algorithm, we calculate the  connections on stable monad bundles defined on the $K3$ twofold and Quintic threefold. An error measure is introduced to determine how closely our algorithmic connection approximates a solution to the Hermitian Yang-Mills equations. We then extend our results by investigating the behavior of non slope-stable bundles. In a variety of examples, it is shown that the failure of these bundles to satisfy the Hermitian Yang-Mills equations, including field-strength singularities, can be accurately reproduced numerically. These results make it possible to numerically determine whether or not a vector bundle is slope-stable, thus providing an important new tool in the exploration of heterotic vacua.
\end{abstract}
  \vspace*{\stretch1}
  \begin{center}
    \texttt{Email: 
      andlara@physics.upenn.edu, 
      vbraun@stp.dias.ie,
      rlk@vt.edu,
      ovrut@elcapitan.hep.upenn.edu
    }
  \end{center}
\end{titlepage}
\tableofcontents

\section{Introduction}
\label{sec:Intro}

\subsection{The Generalized Donaldson Algorithm and Heterotic Compactifications}
A central goal of string theory is to produce low-energy theories with the symmetries, spectrum and properties of elementary particle physics. In the search for realistic vacua, compactification of the heterotic string \cite{Gross:1984dd,Candelas:1985en} and M-theory \cite{Horava:1995qa,Horava:1996ma,Witten:1996mz} on smooth Calabi-Yau threefolds with holomorphic vector bundles \cite{Friedman:1997yq,1997alg.geom..2002D,Lukas:1998hk,Donagi:2000zs,Donagi:1999gc,Blumenhagen:2006ux,Blumenhagen:2006wj} has played an important role \cite{Lukas:1997fg,Lukas:1998yy,Lukas:1998tt,Donagi:1999ez,Donagi:2001fs}.

Heterotic compactifications possess a number of phenomenologically desirable features, including realistic gauge groups and particle spectra \cite{Donagi:2004qk,Braun:2004xv,Donagi:2004ub,Braun:2005ux, Braun:2005zv,Braun:2005nv,Bouchard:2005ag, Anderson:2009mh} as well as  gauge coupling unification \cite{Ambroso:2009sc,Ambroso:2009jd}. However, certain fundamental quantities, such as the Yukawa couplings, are difficult to compute directly \cite{Candelas:1987rx,Candelas:1990rm,Greene:1993vm,Braun:2006me,Donagi:2006yf,Anderson:2009ge}. For example, the ``physical'' Yukawa couplings depend on both the Yukawa coefficients in the
superpotential and the explicit form of the K\"ahler potential. Both quantities are determined by the detailed structure of the underlying geometry--specifically, the metric on the Calabi-Yau threefold and the connection on the slope-stable holomorphic vector bundle--about which, generically, little is known. It follows that the Yukawa couplings in the four-dimensional effective theory have not been explicitly computed, except in very special cases where sophisticated tools from algebraic geometry or topological string theory are available. As a result, one can rarely do better than the qualitative statement that such coupling coefficents either vanish or are ``of order one".

To fully specify the geometry and, hence, to be able to compute the couplings in the effective theory, one must determine the Ricci-flat metric $g$ on the Calabi-Yau threefold $X$ and the supersymmetric gauge connection $A$ on the slope-stable holomorphic vector bundle $\Vsheaf$ \cite{Green:1987sp}. Yau's theorem \cite{MR480350} gives an existence proof for the Ricci-flat metric and the work of Donaldson-Uhlenbeck-Yau \cite{duy1,duy2} provides us with a class of holomorphic vector bundles, called slope-stable, that are consistent with supersymmetry. However, in general the form of $g$ and $A$
are not known analytically. The technical challenge of determining these quantities has made systematic studies of realistic heterotic vacua difficult to achieve. 

In recent years, a number of new numerical approaches to these old problems have been presented \cite{MR2161248,MR1916953, DonaldsonNumerical, MR1064867, MR2154820,Douglas:2006hz, MR2194329,Doran:2007zn,Headrick:2009jz,Douglas:2008es}. Thanks to the development of powerful new algorithms and increased computer speed, it is now possible to compute Ricci-flat metrics and to directly solve the Hermitian Yang-Mills equations for the gauge connection. With this data in hand, all quantities in the four-dimensional effective theory, such as the correctly normalized zero modes, all coefficients in the superpotentials and the K\"ahler potential, can, in principle, be computed. 
In this paper, we take  several important new steps towards this goal. The paper is structured as follows.

In Section \ref{sec:metric}, we outline the general form of Donaldson's algorithm for computing the 
Ricci-flat metric \cite{MR2161248, MR1916953, DonaldsonNumerical} on a Calabi-Yau manifold. Specifically, using the numerical implementations developed in \cite{Braun:2007sn, Braun:2008jp} and \cite{Douglas:2006rr, Douglas:2006hz}, we calculate Ricci-flat metrics on the $K3$ surface and the Quintic threefold. To evaluate the accuracy of our approximation scheme, that is, how close our approximate metric is to the Ricci-flat solution, several measures of the metric-error are introduced.
In Section \ref{sec:HYM}, we turn our attention to recent generalizations of Donaldson's algorithm due to Yau, Wang and Douglas et. al.  \cite{MR2154820,Douglas:2006hz}. These make it possible to compute Hermitian-Einstein metrics on holomorphic vector bundles over a Calabi-Yau manifold and, hence, to solve for the gauge connection satisfying the conditions for ${\cal N}=1$ supersymmetry. 

Given a rank $n$ holomorphic vector bundle $\Vsheaf$ on a Calabi-Yau manifold $X$, the central idea of the algorithm rests on establishing an embedding of $X$ into a Grassmanian $G(n,N_{k}-1)$ via the $N_k$ sections of $\Vsheaf \otimes \cL^{\otimes k}$, where $\cL$ is some ample line bundle.
Using this embedding, a family of simple K\"ahler metrics on $G(n,N_{k}-1)$ can be pulled-back to the Calabi-Yau manifold and used to define a sequence of Hermitian fiber metrics converging to a Hermitian-Einstein metric on $\cal{V}$. In \cite{MR2154820,Douglas:2006rr}, Wang proposed a generalization of Donaldson's ``T-operator" for the embedding described above. It was shown that the iteration of this T-operator converges to a fixed point for each value of $k$ if and only if the bundle $\Vsheaf$ is Gieseker-stable. Furthermore, in the case that $\Vsheaf$ is slope-stable, there exists a scheme of approximate fiber metrics on $\Vsheaf$ which converges to the Hermitian-Einstein metric in the limit that $k \to \infty$. Wang's results are reviewed in Section \ref{sec:generalDonaldson}. We then {\it present a new algorithm to explicitly construct the Hermitian-Einstein fiber metric and, hence, the supersymmetric gauge connection $A$ on $\Vsheaf$}.
An interesting technical aspect arises in the process of extracting the connection on $\cal{V}$ from the connection produced by the algorithm on the ``twisted" bundle $\Vsheaf \otimes \cL^{\otimes k}$. 
In Section \ref{sec:untwist}, by studying all possible methods of ``untwisting" the bundle, we establish the computationally optimal choice and present a general procedure for the computation of the connection on $\Vsheaf$ itself. 
Following Wang's theorem (Theorem \ref{wang1} in Section \ref{sec:generalDonaldson}), we {\it explicitly compute the connection on several slope-stable vector bundles, beginning in Section \ref{sec:error}, and develop an accurate error measure, $\tau_{k_{H}}$, to quantify the accuracy of numerical approximations}.

In Section \ref{sec:StableVsUnstable}, we move beyond Wang's theorem and explore the behavior of non-stable bundles under the generalized Donaldson algorithm. Since Wang's theorem does not apply in this case, there is, a priori,  no reason to expect any specific behavior or systematic results from the algorithm. We show, however, that {\it the output of Donaldson's algorithm, when applied to properly semi-stable or unstable bundles, can in be understood in terms of their Harder-Narasimhan filtrations. Furthermore, it is shown in detail that the failure of these bundles to satisfy the Hermitian Yang-Mills equations, ranging from non-trivial constants to field-strength singularities, can be accurately reproduced numerically}.
These observations demonstrate that Donaldson's algorithm provides an important new tool in the study of the slope-stability properties of vector bundles. In an arbitrary heterotic compactification, one of the most difficult obstacles to overcome is deciding whether or not a given vector bundle is slope-stable. This determination has hithertofore required analytically determining all subsheaves  of the bundle and calculating their slopes. This, at best, is a difficult thing to do, and often cannot be carried out. 
In this paper, we show that {\it  the stability properties of a vector bundle, that is, whether it is stable, semi-stable or unstable, can always be determined by numerical calculation using the generalized Donaldson's algorithm.
This result provides a new tool in the exploration of supersymmetric heterotic vacua. 
We illustrate these procedures by looking at a variety of stable, semi-stable, and unstable bundles on the $K3$ twofold and the Quintic threefold}. The bundles we choose as examples are defined via the monad construction \cite{okonek, Anderson:2007nc, Anderson:2008uw,
  Anderson:2009mh}, a class known to contain bundles corresponding to physically realistic heterotic compactifications. 
  
  In the following two subections, we set the stage for the entire discussion by defining the Hermitian Yang-Mills equations and their relationship to the slope-stabilty of vector bundles \cite{huybrechts}.

\subsection{The Hermitian Yang-Mills Equations and Hermitian-Ein\-stein
Bundle Metrics} 

An $\mathcal{N}=1$ supersymmetric  heterotic string compactification is specified by
a complex three-dimensional Calabi-Yau manifold, $X$, and a holomorphic
vector bundle, $\Vsheaf$, with structure group $K \subset E_{8}$ defined over $X$. The gauge connection, $A$, on $\Vsheaf$ with associated field strength, $F$, must satisfy 
the well-known equations~\cite{Green:1987sp}
\begin{equation}
  \label{the_hym}
  F_{ij}=F_{\bar{i}\bar{j}}=0
  ,\quad
  g^{i\bar{j}}F_{i\bar{j}}=0
\end{equation} 
where $i,j=1,2,3$ run over the holomorphic indices of the Calabi-Yau
threefold. These equations arise as the vanishing condition on the variation of the $10$-dimensional gaugino and, hence, are required to preserve $\mathcal{N}=1$ supersymmetry.
The first two conditions, 
$F_{ij}=F_{\bar{i}\bar{j}}=0$, are simply the constraint that the
vector bundle be holomorphic. On a holomorphic vector bundle,
that is, one with holomorphic transition functions, one
can always choose a connection with a purely $(1,1)$ field strength.
Hence, the vanishing of the $(2,0)$ and $(0,2)$ components is
satisfied. The last condition, $g^{i\bar{j}}F_{i\bar{j}}=0$, however,
is not so easily solved. Equations \eqref{the_hym} are a special
case 
of the Hermitian Yang-Mills
equations. More generally, one has
\begin{equation}
  \label{gen_hym} 
  g^{i\bar{j}}F_{i\bar{j}}
  =
  c \cdot \mathbf{1}
\end{equation} 
where $c$ is a constant determined by the first Chern class of the
bundle $\Vsheaf$. 
For a generic holomorphic, rank $n$ vector bundle, the structure group is
realized in the fundamental representation of $U(n)$. Then $c_1(\Vsheaf)\neq0$ and, therefore, $c \neq 0$.  However, for strucure group $K \subset E_{8}$, it follows that $c_{1}(\Vsheaf)$ must vanish; hence, the zero trace in \eqref{the_hym}.
The numerical algorithms in this paper will require us to find solutions to the 
general equation \eqref{gen_hym}. The results will then be 
specializing to the case $c=0$, that is, \eqref{the_hym}, relevant for
heterotic compactifications.

A solution to \eqref{gen_hym} is equivalent to the
bundle $\Vsheaf$ carrying a particular Hermitian structure. An 
Hermitian structure (or Hermitian fiber metric), $G$, on $\Vsheaf$ is
an Hermitian scalar product $G_{x}$ on each fiber $\Vsheaf(x)$ which
depends differentiably on $x$. The pair $(\Vsheaf,G)$ is often
referred to as an Hermitian vector bundle. For a given choice of frame,
$e_{a}(x)$, for the bundle, one can define the covariant derivative as
\begin{equation} 
  D(v^a e_a)
  =
  (dv^a)e_a + v^a A^{b}_{a}e_{b} \ ,
\end{equation} 
where $a,b=1,\ldots n$ take values in the structure
group
$K \subset U(n)$. The Hermitian
structure defines an inner product as
\begin{equation}
  \label{Gdef} 
  (e_a, e_b)=G_{{\bar a}b}
  ,\quad
  G=G^{\dagger}
  .
\end{equation}
The condition that the connection be compatible with the metric,
\begin{equation} 
  d(u,v)=(Du,v)+(u,Dv)
 \ ,
\end{equation} 
gives 
\begin{equation}
  \partial G_{\bar{a}b}
  =
  G_{\bar{a}c}A^{c}_{b}
\end{equation} 
and, hence,
\begin{equation}
  \label{Amath} 
  A=G^{-1}\partial G
 \ .
\end{equation} 
Since one can always make a gauge choice to set $\bar{A}=0$ for a
holomorphic bundle, we can re-phrase the Hermitian Yang-Mills
equation \eqref{gen_hym} on $F^{(1,1)}$ as a condition on the metric,
\begin{equation}
  \label{herm_met} 
  c \cdot {\bf
    1}
  =
  g^{i\bar{j}}F_{i\bar{j}}
  =
  g^{i\bar{j}}\bar{\partial}_{\bar{j}}A_i
  =
  g^{\bar{j}i}\bar{\partial}_{\bar{j}}(G^{-1}\partial_i G)
\  .
\end{equation} 
A metric $G$ on the fiber of $V$ satisfying the above equation is
called an Hermitian-Einstein metric.

Finally, note that given a fiber metric $G$, one can define an inner product
on the space of global sections, $s^{a}_\alpha$ where $\alpha=1,\ldots h^0(X,\Vsheaf)$, of $\Vsheaf$
 as
\begin{equation}
  \label{sec_prod} 
  <s_{\alpha}|s_{\beta}>
  =\int_X
  \left(s^{b}_{\beta}G_{b\bar{a}}\bar{s}^{\bar{a}}_{\bar{\alpha}}\right)dVol
 \  .
\end{equation}
This will be of use to us in subsequent sections.
Before attempting to explicitly construct a connection $A$ on
$\Vsheaf$ satisfying \eqref{gen_hym} (or, equivalently, an
Hermitian-Einstein bundle metric satisfying \eqref{herm_met}), we will
first rephrase these conditions in terms of algebraic geometry and
recall a central theorem.

\subsection{Slope Stability}
\label{sec:stab} 

Having presented the Hermitian Yang-Mills equations and defined 
Hermitian-Einstein bundle metrics, we now introduce
one more important mathematical notion. 
As stated above, the
Hermitian Yang-Mills equation $g^{i\overline{j}}F_{\overline{j}i}=
c\cdot \bf{1}$ is notoriously difficult to solve for the case
of non-Abelian structure groups.
However, on Calabi-Yau manifolds there exists a powerful method for
transforming this equation into a problem in algebraic geometry. For
K\"ahler manifolds, the Donaldson-Uhlenbeck-Yau
theorem~\cite{duy1,duy2} states that on each \textit{poly-stable}
holomorphic vector bundle $\Vsheaf$ there exists a unique connection
satisfying the Hermitian Yang-Mills equation \eqref{gen_hym}. Thus, to
verify that a vector bundle is consistent with supersymmetry, one need
only verify that it possesses the property of poly-stability.

The notion of slope-stability (also referred to as Mumford-Takemoto
stability) of a bundle (or coherent sheaf), $\cF$, over a K\"ahler
manifold, $X$, is defined by means of a real number called the
\emph{slope}:
\begin{equation}
  \label{slope} 
  \mu (\cF)
  \equiv
  \frac{1}{\rk(\cF)}\int_{X}c_{1}(\cF)\wedge \omega^{d-1} 
  ,
\end{equation} 
where $d$ is the dimension of the K\"ahler manifold. Here,
$\omega$ is the K\"{a}hler form on $X$, while ${\rm rk}(\cF)$ and
$c_1(\cF)$ are the rank and the first Chern class of $\cF$
respectively. A bundle $\Vsheaf$ is called {\it stable} ({\it semi-stable})
if for all sub-sheaves $\cF\subset \Vsheaf$ with $0~<~{\rm rk}(\cF)<~{\rm rk}(\Vsheaf)$ the slope satisfies
\begin{equation}
  \label{slope_req} 
  \mu (\cF) 
  <
  \mu(\Vsheaf)
  \qquad
  (\mu (\cF)\leq \mu (\Vsheaf))
\  .
\end{equation} 
A bundle is {\it poly-stable} if it can be decomposed into a direct sum of
stable bundles 
which all have the same slope. That is, 
%
\begin{equation}
\label{slope_themovie}
\Vsheaf=\bigoplus_n \Vsheaf_n \ , \qquad \mu(\Vsheaf_i)=\mu(\Vsheaf) \ .
\end{equation}
Thus, stability is a
special case of poly-stability which is in turn a subset of
semi-stability. As a series of implications: stable $\Rightarrow$
poly-stable $\Rightarrow$ semi-stable. It is important to note that
the converse to these statements do not hold. For example, not every
semi-stable bundle is poly-stable. Finally, 
observe that the slope is exactly the constant, $c$, that appeared in
\eqref{gen_hym}. Written in terms of the slope of $\Vsheaf$, the
general Hermitian Yang-Mills condition is
\begin{equation}
  \label{hym_genr}
  g^{i\bar{j}}F_{i\bar{j}}
  =
  \mu(\Vsheaf)\cdot \mathbf{1}_{n \times n} \ ,
\end{equation} 
where $n$ is the rank of $\Vsheaf$ and $\mu(\Vsheaf)$ is defined in
\eqref{slope}.

It should be noted that the condition of slope-stability on a
Calabi-Yau manifold depends on all moduli of the heterotic
compactification. To be specific, consider a
Calabi-Yau threefold. Here, the moduli are the K\"ahler
moduli ($h^{1,1}(X)$), the complex structure moduli ($h^{2,1}(X)$) and
the vector bundle moduli ($h^1(End(\Vsheaf))$). The dependence on K\"ahler
moduli is explicit in \eqref{the_hym} and referred to as a choice
of ``polarization''. One can expand the
K\"ahler form $\omega$ in \eqref{slope} as
$\omega=t^{r}\omega_{r}$, where $\omega_{r}$ form a basis of 
$(1,1)$-forms and $t^r$ are the real parts of the K\"ahler moduli. Written in
terms of the triple intersection numbers, $d_{rst}$, of the threefold,
the slope is simply $\mu(\Vsheaf)=\frac{1}{rk(\Vsheaf)}d_{rst}
c_1(\Vsheaf)^r t^s t^t$ where $r,s=1,\ldots h^{1,1}(X)$. The other
moduli enter through the notion of a subsheaf $\cF \in \Vsheaf$. In
general, whether or not there exists an injective sheaf morphism
$\phi: \cF \to \Vsheaf$ depends on both the complex structure moduli
and the vector bundle moduli.

Thus, finding a solution to the Hermitian Yang-Mills equations (that
is, determining whether the bundle is slope-stable) is a question that
must be asked after selecting a particular point in moduli space. As
we will see, some regions of moduli space may admit a solution while
others will not. This moduli dependence can lead to a variety of interesting
physical consequences~\cite{Sharpe:1998zu, Anderson:2009sw,
  Anderson:2009nt, Anderson:2010tc}, which we will explore in later
sections. For most of this work, however, we will be interested in
studying the stability properties of bundles at ``generic'' points in
their moduli space.

Before investigating solutions to the Hermitian Yang-Mills equations,
we must first introduce an algorithmic approach to determining
Hermitian metrics (both of manifolds and bundles). In the following
section, we discuss Donaldson's algorithm for numerically
approximating the Ricci-flat metric on a Calabi-Yau manifold.


\section{Computing the Calabi-Yau Metric}
\label{sec:metric}

\subsection{Donaldson's Algorithm}
\label{sec:donaldson}

Many of the challenges associated with string compactifications on a
Calabi-Yau threefold, $X$, arise from the difficulty in
determining the explicit geometry. The simplest ${\cal{N}}=1$ supersymmetric vacuum solutions require a Ricci-flat metric, $g_{i\bar{j}}$,
on $X$. While Yau's theorem~\cite{MR480350} ensures that such a
metric exists, no analytic solutions have been found. Recently, however, it has become possible to find numerical solutions using an algorithm developed by
Donaldson~\cite{MR2161248, MR1916953, DonaldsonNumerical}. This
algorithm has been implemented numerically and extended
in~\cite{MR2283416, Braun:2007sn, Braun:2008jp, Douglas:2006rr,
  Douglas:2006hz, Headrick:2009jz}. For other numerical computations of
K\"ahler metrics, see~\cite{MR2194329, Headrick:2005ch, Doran:2007zn}.
In this paper, we propose to investigate solutions to the
Hermitian Yang-Mills equation $g^{i\bar{j}}F_{i\bar{j}}=0$. Since the Ricci-flat metric,
$g_{i\bar{j}}$, is required for such a solution, in this
section we provide a brief review of Donaldson's
algorithm for approximating such metrics
on Calabi-Yau manifolds. 

The starting point of Donaldson's algorithm is the observation that
there exists a natural metric on 
$\mathbb{P}^n$. Denoting the $n+1$ homogeneous coordinates by $z_{i}$, this Fubini-Study metric is given by $g_{FS
  i\bar{j}}=\frac{i}{2}\partial_{i} \bar{\partial}_{\bar{j}}K_{FS}$,
where
\begin{equation}
  \label{FS} 
  K_{FS} = \frac{1}{\pi}{ \rm{ln}} \sum_{i\bar{j}}
  h^{i\bar{j}}z_{i}\bar{z}_{\bar{j}}  
\end{equation} 
and $h^{i\bar{j}}$ is any hermitian, non-singular matrix. 
The Fubini-Study metric is usually defined with
$h^{i\bar{j}}=\delta^{i\bar{j}}$, but we will refer to \eqref{FS} as a
Fubini-Study K\"ahler potential for any hermitian matrix $h^{i\bar{j}}$. 
The Fubini-Study metric can be used
to induce a metric on any subvariety of $\mathbb{P}^n$. In particular, it will induce a metric on any Calabi-Yau threefold $X$, since
it is always possible to embed $X \subset \mathbb{P}^n$ for some
$n$.
However, such a metric will not, in general, be Ricci-flat.
To obtain the Ricci-flat metric, Donaldson's algorithm requires one to 
use a generalized version of \eqref{FS} with enough
free parameters to provide a more versatile induced metric on $X$.  The algorithm then
guarantees that this generalized metric will converge, in some specified limit, to the Ricci-flat metric.

An obvious generalization of \eqref{FS} is to replace the
degree one polynomials with polynomials of higher
degree. For example,
\begin{equation}
  \label{FS_gen} 
  K = \frac{1}{k\pi} {\rm{ln}} \sum_{i_{1}\ldots
    i_{k}\bar{j}_{1}\ldots\bar{j}_{k}} h^{i_{1}\ldots
    i_{k}\bar{j}_{1}\ldots\bar{j}_{k}}z_{i_{1}}\ldots
  z_{i_{k}}\bar{z}_{\bar{j}_{1}}\ldots \bar{z}_{\bar{j}_{k}}
\end{equation} 
where $h^{i_{1}\ldots i_{k}\bar{j}_{1}\ldots\bar{j}_{k}}$ is hermitian.
This new K\"ahler potential now has $(n+1)^{2k}$ real parameters. 
One can write the above generalization in a more systematic way by
noting that such K\"ahler potentials can naturally be obtained by considering holomorphic
line bundles on the Calabi-Yau manifold $X$ itself and using the Kodaira embedding theorem. 
Let $\Lsheaf$ be a holomorphic line bundle over $X$ with $n_{1}=h^0(X,
\Lsheaf)$ global sections. Consider a twisting of the line bundle,
$\mathcal{L}^{k} =\Lsheaf^{\otimes k}$. Then, choosing a basis for the
space of sections, $s_{\alpha} \in H^0(X,\Lsheaf^k)$ where $0
\leq \alpha \leq n_{k}-1$, allows one to define a map from $X$ to
$\mathbb{P}^{n_{k}-1}$ given by
\begin{equation}
  \label{embed} 
  i_{k}:~ X \to \mathbb{P}^{n_{k}-1} 
  ,\quad
  (x_{0},\ldots,x_{2}) \mapsto \big[ s_0(x): \ldots : s_{n_{k}-1}(x) \big]
  \ ,
\end{equation} 
where $x_i$ are holomorphic coordinates on the Calabi-Yau threefold. The Kodaira embedding theorem~\cite{MR507725} states that if
$\mathcal{L}$ is ample, \eqref{embed} will define an embedding of $ X
\subset \mathbb{P}^{n_{k}-1}$ for all $\mathcal{L}^k$ with $k \geq
k_0$ for some $k_0$.

In terms of this natural embedding line bundle, we can view the
generalized K\"ahler potential \eqref{FS_gen} restricted to $X$ simply as
\begin{equation}
  \label{Lmetric} 
  K_{h,k}
  =\frac{1}{k\pi} \ln \sum_{\alpha,\bar{\beta}=0}^{n_{k}-1} h^{\alpha\bar{\beta}}
  s_{\alpha}{\bar{s}}_{\bar{\beta}}=\ln ||s||^{2}_{h,k} \ .
\end{equation} 
Geometrically, \eqref{Lmetric} defines an hermitian
metric on the line bundle $\Lsheaf^{k}$ itself. It provides a natural
inner product on the space of global sections
\begin{equation}
  \label{sec_metric}
  M_{\alpha\bar{\beta}}=\left<s_{\beta}|s_{\alpha}\right> 
  = 
  \frac{n_{k}}{\Vol_{CY}(X)} \int_{X}
  \frac{s_{\alpha}{\bar{s}}_{\bar{\beta}}}{||s||^{2}_{h}}\dVol_{CY} \ ,
\end{equation} 
where
\begin{equation}
\label{metric_themovie}
{\rm dVol_{CY}}=\Omega \wedge {\bar{\Omega}}
\end{equation}
and $\Omega$ is the holomorphic (3,0) volume form on $X$. Note that \eqref{sec_metric} depends non-linearly on $h$.
It was shown by Tian~\cite{MR1064867} that
metrics on ample line bundles, such as \eqref{Lmetric}, can provide a
useful ``basis" of Kahler metrics on $X$. That is, any algebraic
K\"ahler potential can either be written in the form \eqref{Lmetric}
or obtained as the limit point of a sequence of such potentials. Specifically, one has the following theorem.
\begin{theorem}[Tian] 
  Let $\{s_\alpha \}$ be a basis for $H^0(X,\Lsheaf^k)$ for some
  ample line bundle $\Lsheaf$. Then the space of all ``algebraic''
  \Kahler{} potentials,
  \begin{equation}\label{k_basis} K_{h,k} =\frac{1}{k\pi} \ln \sum_{\alpha,\bar
      \beta=0}^{n_{k}-1} h^{\alpha\bar\beta} s_\alpha \sbar_{\bar\beta}
  \end{equation} where $k\in\Z$, is dense in the space of \Kahler{}
  potentials.
\end{theorem}

With this observation, we return to the goal of finding the Ricci-flat
metric on $X$. Consider $X$ as a projective manifold, with an ample
line bundle $\Lsheaf$. Note that, in general, the matrices
$h^{\alpha\bar{\beta}}$ and $M_{\alpha\bar{\beta}}$ in
\eqref{sec_metric} are completely unrelated. However, for special metrics, they may coincide.
The metric $h$ on the line bundle $\Lsheaf$ is called {\it ``balanced"} if
\begin{equation} 
  (M_{\alpha\bar{\beta}})^{-1}
  =
  h^{\alpha\bar{\beta}} 
  .
\end{equation} 
In this case, one can find an ``orthonormal'' basis of sections for
which $M_{\alpha\bar{\beta}}=\delta_{\alpha\bar{\beta}}$ and
$h^{\alpha\bar{\beta}}=\delta^{\alpha\bar{\beta}}$. Many theorems have
been proven about balanced metrics. Of particular interest here is
their curvature properties~\cite{MR2161248, MR1916953,
  DonaldsonNumerical}.

\begin{theorem}[Donaldson] 
  For each $k \geq 1$, the balanced metric, $h$, on $\Lsheaf^k$ exists
  and is unique. As $k \to \infty$, the sequence of metrics
  \begin{equation}
    \label{g_approx}
    g_{i\bar{j}}^{(k)}
    =
    \frac{1}{k\pi}\partial_{i} \bar{\partial}_{\bar{j}}
    \ln \sum_{\alpha,\bar{\beta}=0}^{n_{k}-1}
    h^{\alpha\bar{\beta}}s_{\alpha}\bar{s}_{\bar{\beta}}
  \end{equation} 
  on $X$ converges to the unique Ricci-flat metric for the given
  K\"ahler class and complex structure.
\end{theorem}

The core of Donaldson's algorithm is then simply the task of finding
the balanced metric for each $k$. To this end, Donaldson defines the
``T-operator"
\begin{equation}
  \label{t_oper}
  T(h)_{\alpha\bar{\beta}}
  =
  \frac{n_{k}}{\Vol_{CY}(X)}
  \int_{X}
  \frac{s_{\alpha}{\bar{s}}_{\bar{\beta}}}{\sum_{\gamma\bar{\delta}}
    h^{\gamma\bar{\delta}}s_{\gamma}{\bar{s}}_{\bar{\delta}}}\dVol_{CY}
\end{equation} 
which, for a metric $h$, computes the matrix $T(h)$. Given
a fixed point of this operator, that is, a metric $h$ for which $T(h)=h$, one can always
perform a change of basis $s \to h^{-1/2}s$ to bring $h$ to the unit
matrix, producing a balanced embedding. To find this fixed point,
simply iterate \eqref{t_oper} as follows.
\begin{theorem}[Donaldson] 
  For any initial metric $h_0$ (and basis $s_{\alpha}$ of
  global sections of $\Lsheaf^k$), the sequence
  \begin{equation} 
    h_{n+1}=(T(h_n))^{-1}
  \end{equation} 
  converges to the balanced metric as $n \to \infty$.
\end{theorem} 
\noindent Happily, in practice, very few $(\leq 10)$ iterations are needed to
approach the fixed point. Henceforth, we will also refer to  $g^{(k)}_{i\bar{j}}$ in \eqref{g_approx}, the approximating metric for fixed $k$, as a
balanced metric.

To find the balanced metric at each step $k$, one must be able
to integrate over the Calabi-Yau threefold. An explicit numerical
integration scheme for this purpose was given in~\cite{Braun:2007sn,
  Braun:2008jp} and is used to compute the Calabi-Yau
metrics in this paper.  We refer the reader to these references for
details of this and other technical aspects of the computer
implementation.  To summarize Donaldson's algorithm:

\begin{enumerate}
\item Choose an ample line bundle $\mathcal{L}$ and a degree $k$ (that
  is, a twisting of the line bundle $\mathcal{L}^k$) at which to
  compute the balanced metric which will approximate the
  Calabi-Yau metric.
\item Calculate a basis $\{s_{\alpha}\}_{\alpha=0}^{n_{k}-1}$ for
  $H^0(X,\mathcal{L}^{k})$ at the chosen $k$.
\item Choose an initial non-singular, hermitian matrix,
  $h^{\gamma\bar{\delta}}$. Perform the numerical integration to compute
  the T-operator in \eqref{t_oper}.
\item Set the new $h^{\alpha\bar{\beta}}$ to be
  $h^{\alpha\bar{\beta}}=(T_{\alpha\bar{\beta}})^{-1}$.
\item Return to item $3$ and repeat until $h^{\alpha\bar{\beta}}$
  approaches its fixed point. In practice, this convergence occurs in
  less than $10$ iterations and does not depend on the initial choice of
  $h^{\alpha\bar{\beta}}$.
\end{enumerate}
We now turn our attention to how one can measure convergence in
Donaldson's algorithm. That is, for each finite step $k$, how does one
define the error measure which tells us how close we are to the unique
Ricci-flat metric?

\subsection{Ricci Curvature and Error Measures}
\label{sec:ricci}

We begin by defining the Calabi-Yau volume form and several other
useful quantities. Let $X$ be a smooth Calabi-Yau variety of dimension
$d$, and let $\Omega$ denote the unique (up to a constant) $(d,0)$
form. The associated volume is given by
\begin{equation}
  \label{vol} 
  \Vol_{CY} = \int_X \Omega \wedge \Omegabar
\  .
\end{equation} 
Given an ample line bundle $\mathcal{L}$, let $\omega_k$ denote the
Kahler form
\begin{equation}
  \label{kform}
  \omega_k
  =
  \frac{i}{2}g^{(k)}_{i\bar{j}}dz_{i}\wedge d\bar{z}_{\bar j}
\end{equation} 
of the balanced metric associated with $\mathcal{L}^k$. Note that the
K\"ahler class associated with this K\"ahler form is $[\omega_k]=2\pi
c_{1}(\mathcal{L}^k)$. The associated volume is
\begin{equation} 
  \Vol_k = \frac{1}{d!}\int_X \omega_k^d
\  ,
\end{equation} 
where $\omega_{k}^{d}$ denotes the $(d,d)$ volume form $\omega \wedge \dots \wedge \omega$.
As expected, when numerically computed, this value is close to $ k^d
\int_X \omega_1^d$. If, for example, we restrict to the case of the
degree $d+2$ hypersurface in $\CP^{d+1}$, then by Poincare duality we
have that
\begin{equation} 
  \int_X \omega_1^d 
  = 
  (d+2) \int_{\CP^{d+1}}
  \left(\omega^{\CP^{d+1}}_1\right)^d
  ,
\end{equation} 
which allows us to fix the overall normalization to $\int_X \omega_1^d
= d+2$.

These two volumes allow one to define several measures of the 
convergence of the balanced metric to the Ricci-flat metric. The first is $\sigma_k$ introduced
in~\cite{Braun:2007sn, Braun:2008jp} and given by
\begin{equation}
  \label{e:sig} 
  \sigma_k = \frac{1}{\Vol_{CY}}
  \int_X \left| 1-
    \frac{\omega_k^d / \Vol_k  }{\Omega \wedge \Omegabar  / \Vol_{CY}} 
   \right| \dVol_{CY} \ .
\end{equation} 
Note that the K\"ahler form in \eqref{kform} is the Calabi-Yau
K\"ahler form if and only if $\omega_{k}^3$ is
proportional to $\Omega \wedge \Omegabar $. Since we
know the exact volume form $\Omega\wedge\bar{\Omega}$, only $\omega_{k}$
is approximate in \eqref{e:sig} and depends on the degree of
$k$. Hence, the integral in \eqref{e:sig} vanishes if and only if
$\omega_{k}$ is the Calabi-Yau K\"ahler form. As $k$ is increased in
Donaldson's algorithm, $\sigma_k$ should approach
zero. In~\cite{Braun:2007sn, Braun:2008jp}, there is a prediction as
to how $\sigma_k$ should converge to zero. They find that the error
approaches zero at least as rapidly as
\begin{equation}
  \label{sigma_decay}
  \sigma_k=\frac{a_2}{k^{2}}+\frac{a_3}{k^3}+\ldots
\end{equation} 
for some constants $a_i$. This prediction is verified by our results
in the next subsection.

The second error measure we use is a global measure of convergence for
the Ricci scalar,
\begin{equation}
  \label{e:R} 
  ||R||_k = 
  \frac{\Vol_k^{1/d} }{ \Vol_{CY} } 
  \int |R_k|\, \dVol_{k}
\  .
\end{equation} 
$R_k$ is the Ricci scalar computed with the balanced metric $\omega_k$
and we integrate its absolute value. Note that by including the factor
$\Vol_k^{1/d} $, the naive linear $k$ dependence of the integral
inherited from the fact that $c_1(\mathcal{L}^k)=kc_1(\mathcal{L})$ is
canceled. As the balanced metrics converge to the Ricci-flat metric for increasing $k$, this measure should approach zero.

Third, instead of using the exact Calabi-Yau volume form, one can also use
the volume-form computed from the approximate Calabi-Yau metric. This
is nothing but the Einstein-Hilbert action, so we define
\begin{equation}
   EH_k =\Vol_k^{(1-d)/d} 
    \int R_{k} \; \frac{\omega_{k}^d}{d!}  
    =
    \Vol_k^{(1-d)/d} \int R_{k} \; \sqrt{\det g_{k}} 
    \; \text{d}^{2d}\! x \ .
\label{again1}
\end{equation}
Note, however, that on a \Kahler{} manifold the
closed two-form $\Ric = R_{i \bar j}\diff{}z^i \diff{}\bar z^{\bar j}$
defines the first Chern class of $X$, since $c_1(X)=[\frac{Ric}{2\pi i}]$. Furthermore, $\Ric \wedge \omega^{d-1} = 2
d R \omega^d$ and hence, the Einstein-Hilbert action
\begin{equation}\label{zero_test} 
  EH_k =  \Vol_k^{(1-d)/d} \int_X R_k \; \omega^d =
 \Vol_k^{(1-d)/d} \int_X c_1(X) \wedge \frac{\omega^{d-1}}{d!} 
  =
  0
\end{equation} 
vanishes on a Calabi-Yau manifold for any metric (and hence for any integer $k$). As a result, we can use this to directly test the accuracy of our numerical integration. However, a better measure of the convergence of the balanced metrics to the Ricci-flat metric is given by
\begin{equation}
   ||EH||_k = 
   \Vol_k^{(1-d)/d} 
    \int |R_{k}| \; \sqrt{\det g_{k}} \; \text{d}^{2d}\! x 
\    .
\label{again2}
\end{equation}
On a Calabi-Yau manifold, $||R||_k, ||EH||_k =
O(k^{-1})$ as $k\to \infty$ and, hence, these error measures should approach zero.

\subsection{ The Quartic $K3$ }
\label{sec:QuarticK3}

In the next two subsections, we use the algorithm outlined above to
numerically compute the Ricci-flat metric on two Calabi-Yau
manifolds. Two separate computer implementations of Donaldson's
algorithm were developed in~\cite{Braun:2007sn, Braun:2008jp}
and~\cite{Douglas:2006rr, Douglas:2006hz}. We have utilized these two
independent sets of code to check the accuracy of our results. These
will be denoted as \texttt{Code1} and \texttt{Code2} respectively in the following.

Begin with the simple case of the $K3$ twofold. In this subsection, we
consider the one parameter deformation $X$ of the Fermat quartic in
$\CP^3$ given by
\begin{equation}
  \label{k3} 
  X\colon \quad 
  \sum_{i=0}^3 z_i^4 - 4 \psi \prod_{i=0}^3 z_i 
  = 0 
  .
\end{equation} 
Here $z_0,\ldots z_3$ are the homogeneous coordinates on $\CP^3$. The deformation parameter $\psi$ is, on the face of it, a complex
number. However, redefining 
\begin{equation}
  \psi \mapsto e^{2 \pi i / 4} \psi
  ,\quad 
  \big[ z_0: z_1: z_2: z_3 \big] 
  \mapsto
  \left[ \mapsto e^{-2 \pi i / 4} z_0: z_1: z_2: z_3 \right] 
\end{equation}
leaves $X$ invariant, leading to a $\Z_4$ isometry acting on the naive
modulus $\psi$. In other words, the actual modulus is $\psi^4$, and
the complex structure moduli space has an orbifold singularity at the
Fermat point $\psi=0$. As is well known, the algebraic variety $X$ is
smooth as long as $\psi$ is not a fourth root of unity and is away from
the large complex structure limit, that is, $\psi^4\neq1, \infty$.

\begin{figure}
  \begin{center} \input{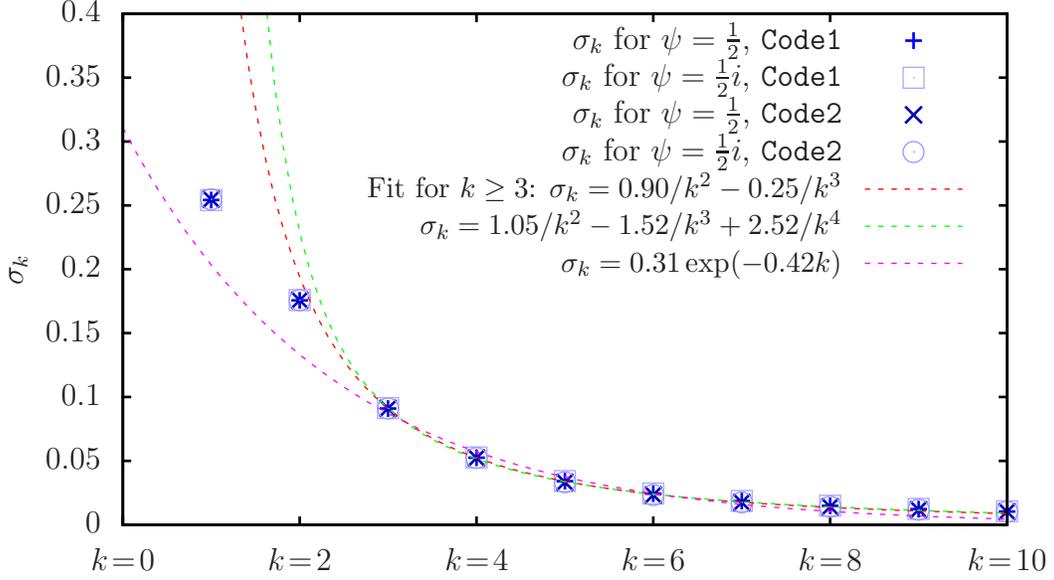}
  \end{center}
  \caption{The error measures $\sigma_k$ defined in Subsection
    \ref{sec:ricci}. The data shown is for the Quartic $K3$ defined as a
    hypersurface in $\mathbb{P}^3$ \eqref{k3}. The complex structure
    parameter is chosen to be $\psi=\frac{1}{2}$ and
    $\psi=\frac{i}{2}$. Shown is data generated using the code developed
    in~\cite{Braun:2007sn, Braun:2008jp} (\texttt{Code1}) and data
    generated by the implementation in~\cite{Douglas:2006rr,
      Douglas:2006hz} (\texttt{Code2}). The error measure is fitted to the
    theoretical convergence given in \eqref{sigma_decay}.}
  \label{fig:MetricError-K3-sigma}
\end{figure}
\begin{figure}[p]
  \begin{center} \input{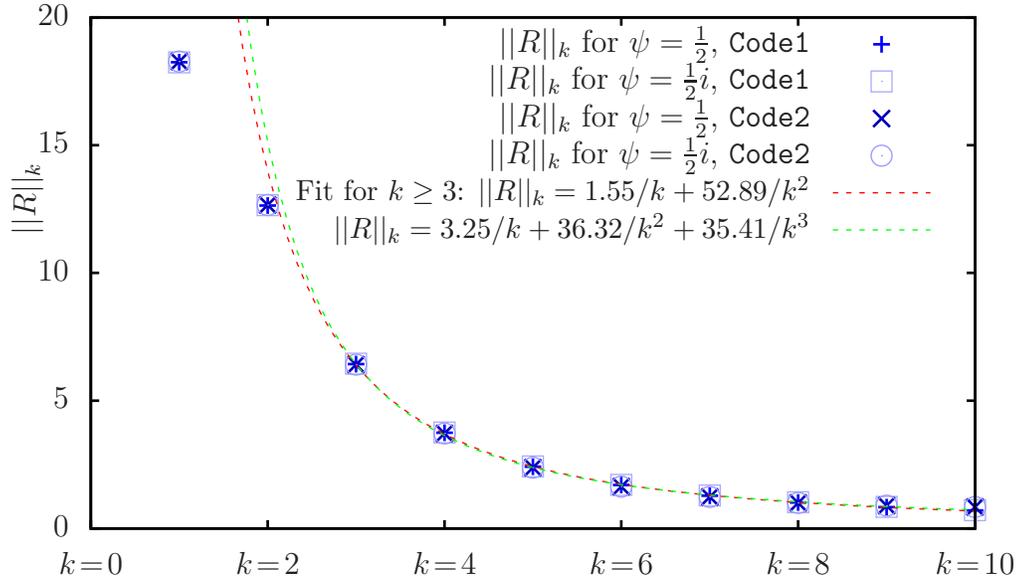}
  \end{center}
  \caption{Average scalar curvature for the metric on $K3 \subset
    \CP^3$.}
  \label{fig:MetricError-K3-curvature}
\end{figure}
\begin{figure}[p]
  \begin{center} \input{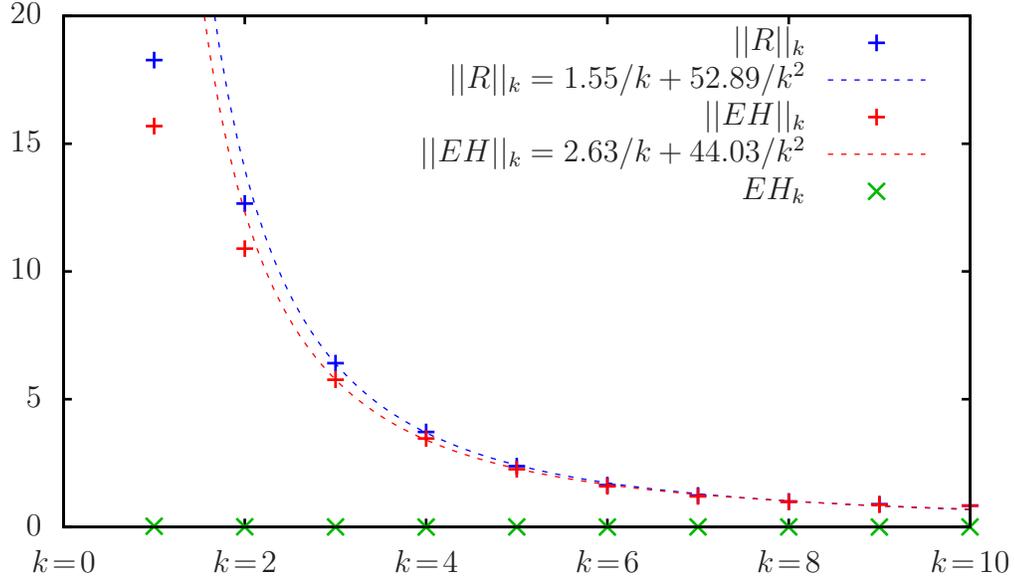}
  \end{center}
  \caption{Curvature measures for $K3 \subset \CP^3$. These provide
    the additional checks of numeric accuracy described in
    \autoref{sec:ricci}.}
  \label{fig:MetricError-K3-EH}
\end{figure}
The very ample line bundle we consider is
$\mathcal{L}=\Osheaf_X(1)$. For each $k$, we computed the quantities
in question for both $\psi=0.5$ and $\psi=0.5 e^{2 \pi i / 4}$. These
two parameter values are a significant check of the validity of our
numerical implementation since these two $K3$ metrics are, in fact,
identified under the $\mathbb{Z}_4$ discrete isometry of $X$. We would
expect the metric results for these two values of $\psi$ to be
identical up to numerical errors. This is confirmed by the data shown
in Figures~\ref{fig:MetricError-K3-sigma},
\ref{fig:MetricError-K3-curvature}, and~\ref{fig:MetricError-K3-EH}.

In these plots, the T-map was iterated with \comma{1600000} points,
and the error measures were computed with \comma{500000} points. The
high accuracy allows one to conclude that both code bases give the same
result. Moreover, especially for high $k$ the assumption that
$\sigma_k=O(\tfrac{1}{k})$ fits our data better than exponential
fall-off, see~\cite{Braun:2008jp, Headrick:2009jz}.

\subsection{The Quintic Threefold}
\label{eq:quintic}

Now consider a Calabi-Yau threefold; specifically, the one parameter deformation $X$ of the Fermat quintic in
$\CP^4$ given by
\begin{equation}\label{quintic} X\colon \sum_{i=0}^4 z_i^5 - 5\psi
  \prod_{i=0}^4 z_i = 0 \ .
\end{equation} The algebraic variety $X$ is smooth as long as $\psi$
is not a fifth root of unity.  
The T-map was iterated with
\comma{2000000} points, and the error measures were 
 computed with
\comma{500000} points.

\begin{figure}
  \begin{center} \input{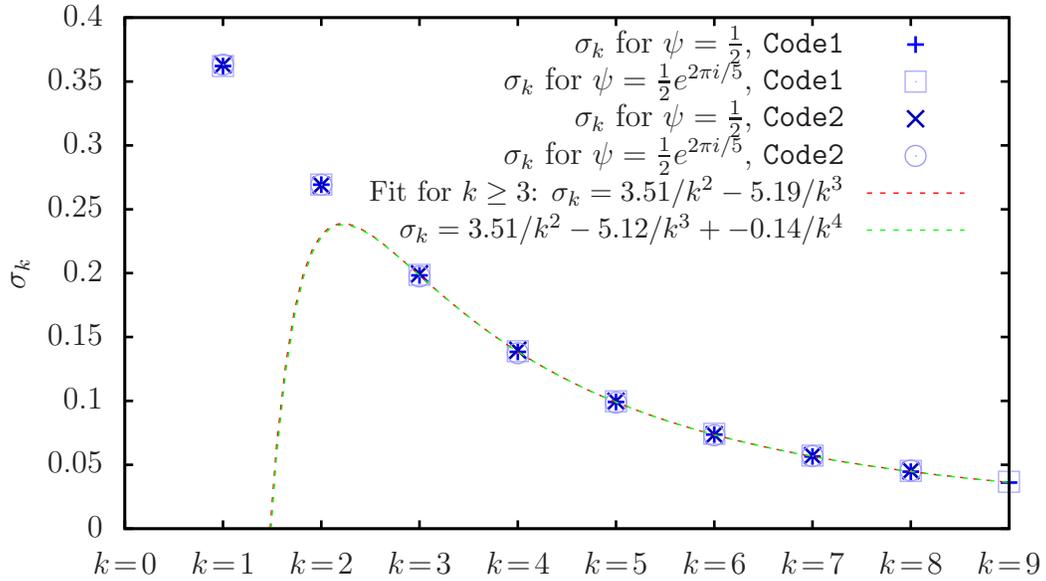}
  \end{center}
  \caption{The $\sigma_k$ error measures for the quintic threefold
    \eqref{quintic} in $\mathbb{P}^4$. Shown is the error measure
    described in Subsection \ref{sec:ricci}, evaluated for the two values
    $\psi=\frac{1}{2}$ and $\psi=\frac{i}{2}$. \texttt{Code1} and
    \texttt{Code2} are associated to the implementations
    of~\cite{Braun:2007sn, Braun:2008jp} and~\cite{Douglas:2006rr,
      Douglas:2006hz} respectively.}
  \label{fig:MetricError-Quintic-sigma}
\end{figure}

\begin{figure}
  \begin{center} \input{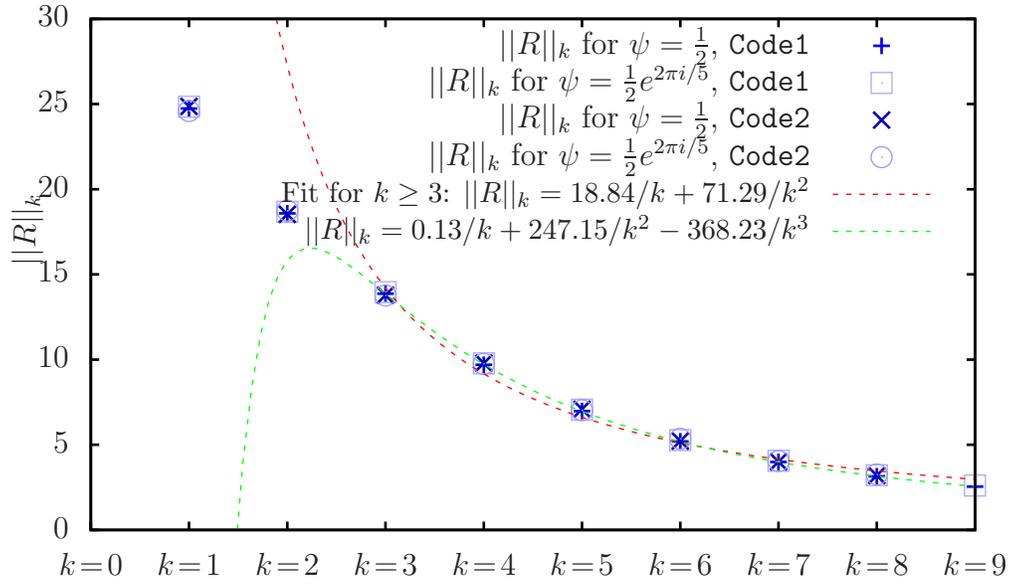}
  \end{center}
  \caption{Average scalar curvature for the metric on the Quintic,
    \eqref{quintic}.}
  \label{fig:MetricError-Quintic-curvature}
\end{figure}

\begin{figure}
  \begin{center} \input{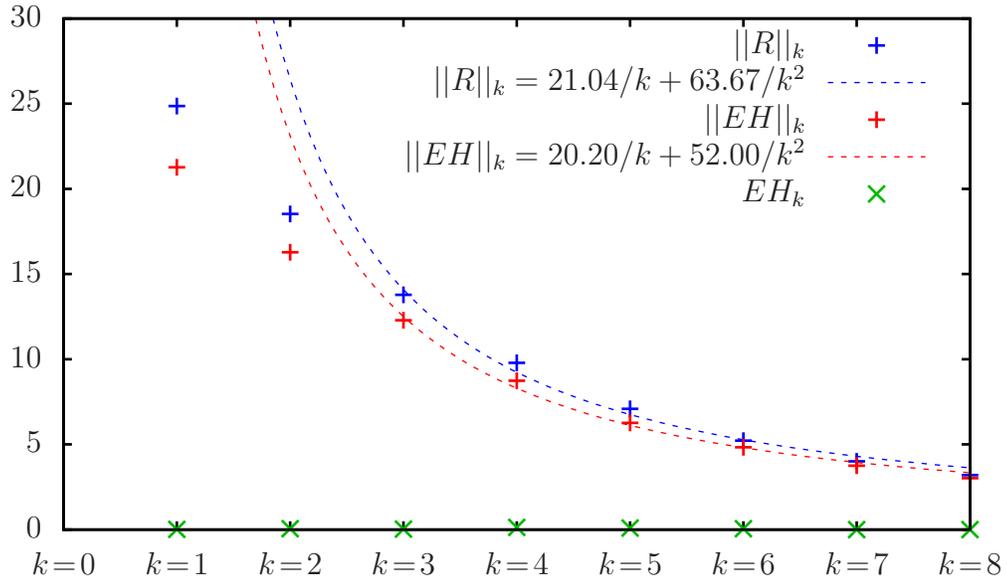}
  \end{center}
  \caption{Curvature measures for the Quintic, \eqref{quintic}.}
  \label{fig:MetricError-Quintic-EH}
\end{figure}

The very ample line bundle we consider is again
$\mathcal{L}=\Osheaf_X(1)$. For each $k$, we computed the quantities
in question for both $\psi=0.5$ and $\psi=0.5 e^{2 \pi i / 4}$. As
described in the previous subsection, due to the discrete
$\mathbb{Z}_{5}$ isometry of the quintic given in \eqref{quintic}, we
would expect the metric results to be identical for these two
choices. Once again, this is verified by the plots in
\autoref{fig:MetricError-Quintic-sigma}
and~\ref{fig:MetricError-Quintic-curvature}.
 We also repeated the computation of $||EH||_k$ compared
to $||R||_k$.  The result is shown in
\autoref{fig:MetricError-Quintic-EH}.
As discussed in the Subsection \ref{sec:ricci}, using the identity in \eqref{zero_test} one can estimate the accuracy of the numerical integration from
$EH_k\approx 0$.


\section{Hermitian Yang-Mills Connections}
\label{sec:HYM}

\subsection{Generalizing Donaldson's Algorithm}
\label{sec:generalDonaldson}

As we saw in the previous section, Donaldson's algorithm is a powerful
tool for numerically approximating the Calabi-Yau metric. In this
section, we investigate a generalization of these techniques
which can be used to approximate the field strength, $F^{(1,1)}$, of a
holomorphic connection which satisfies \eqref{hym_genr}. As
discussed in \autoref{sec:metric}, Donaldson's algorithm can be viewed
as a method for numerically obtaining a particular Hermitian structure
on the ample line bundle $\cL^k$, see \eqref{sec_metric}. This
``balanced'' fiber metric on $\cL^k$ allows one to define a balanced
embedding of the Calabi-Yau space $X$ into $\mathbb{P}^{n_{k}-1}$.
By mapping the coordinates $x \in X$ into the global sections
$s_{\alpha} \in H^0(X,\cL^{k})$, that is,
\begin{equation}
  \xymatrix@R=5mm@C=2cm{
    x \ar@{|->}[r] & 
    [ s_0(x) : \cdots : s_{n_k-1}(x)] \ ,
  }
\end{equation}
we produce a map $i_{k}: X \to \mathbb{P}^{n_{k}-1}=G(1,n_{k}-1)$ where
$n_k=h^0(X,\cL^k)$. The pull-back of the associated Fubini-Study
metric was shown in \autoref{sec:donaldson} to converge to the
Ricci-flat metric on $X$ in the limit that $k \to \infty$. It is a
natural question to ask whether or not an analogous algorithm could be
developed to approximate Hermitian metrics on higher rank vector
bundles? In particular, could one find an approximation scheme to
produce an Hermitian metric on an arbitrary stable bundle $\Vsheaf$ of
rank $n$ such that it satisfies condition \eqref{herm_met}?
Fortunately, precisely this question has been addressed in the
mathematics literature~\cite{MR2154820} and in~\cite{Douglas:2006hz}.

To begin a generalization of Donaldson's algorithm, 
consider defining an embedding via the global sections of a twist of
some holomorphic vector bundle, $\Vsheaf$, with non-Abelian structure
group. That is, consider a map
\begin{equation}
  \xymatrix@R=5mm@C=2cm{
    x \ar@{|->}[r] & 
    {
      \left[
        \begin{pmatrix}
          S_1^1(x) \\ \vdots \\ S_1^n(x)     
        \end{pmatrix}
        : \cdots :
        \begin{pmatrix}
          S_{N_k}^1(x) \\ \vdots \\ S_{N_k}^n(x)
        \end{pmatrix}
      \right]
      .
    }}
\end{equation}
from the coordinates, $x$, of $X$ into the global sections $S^{a}_{\alpha}
\in H^0(X,\Vsheaf\otimes\cL^k)$, where $\alpha=0 \ldots
N_k-1=h^0(X,\Vsheaf\otimes \cL^k)$ (the number of global sections of
$\Vsheaf\otimes \cL^k$) and the index $a=1,\ldots n$ is valued in
the fundamental representation of structure group $K \subset U(n)$ of
the rank $n$ bundle $\Vsheaf$. We hope then to define the embedding
\begin{equation}
  \label{V_embed}
  X \rightarrow G(n,N_{k}-1) \ ,
\end{equation}
where $G(n,N_{k}-1)$ denotes the Grassmannian
of the relevant dimension. By the Kodaira embedding
theorem~\cite{MR507725}, given a holomorphic
vector bundle $\Vsheaf$ and an ample line bundle $\cL$, there
must exist a finite integer $k_0$ such that, for any $k>k_0$, a ``twist'' of
the bundle $\Vsheaf (k)=\Vsheaf\otimes \cL^k$ defines an embedding,
$i_k: X \to G(n,N_k-1)$.

Such twisting is a clear necessary for the bundles of
interest in $E_8 \times E_8$ heterotic
compactifications. Because a slope-stable bundle $\Vsheaf$ in a
heterotic compactification cannot have any global sections\footnote{To see why this is the case, note that for the structure group $K$ of $\Vsheaf$ to embed inside $E_8$ one must require that $K=SU(n)$ for $n=3,4,5$ and, hence, that
  $c_1(\Vsheaf)=0$. Therefore, from \eqref{slope} it follows that
  $\mu(\Vsheaf)=0$. However, if $H^0(X,\Vsheaf) \neq 0$ then $\cO_X$ must inject into $\Vsheaf$ in contradiction with
  the assumption of stability.},
$H^0(X,\Vsheaf)=0$, we begin by twisting the bundle
by some sufficiently large power of an ample line
bundle $\cL$. That is, we consider $\Vsheaf \otimes \cL^k$. If $\cL$
is ample, then $\Vsheaf \otimes \cL^k$ will be
generated by its global sections; that is, it will define an embedding as
in \eqref{V_embed}. In our search for a solution to the Hermitian
Yang-Mills equation \eqref{hym_genr}, the connection on the twisted
bundle will be closely related to the original connection, since such a
twist only modifies the trace part of the field strength. Stated in
terms of algebraic geometry, the process of twisting will not modify
the slope-stability properties of $\Vsheaf$ since $\Vsheaf\otimes
\cL^k$ is stable if and only if $\Vsheaf$ is.

As at the beginning of \autoref{sec:donaldson}, where we chose the trial
form of the K\"ahler potential in \eqref{Lmetric}, here we begin with
another simple anzatz; this time, however, for the
Hermitian structure $G$ in \eqref{Gdef}. We consider an Hermitian
matrix $(G^{-1})^{a\bar{b}}$ of the form
\begin{equation}
  \label{G_anzatz}
  (G^{-1})^{a\bar{b}}
  =
  \sum_{\alpha,\beta=0}^{N_k-1}H^{\alpha\bar{\beta}}S_{\alpha}^{a}
  (\bar{S})_{\bar{\beta}}^{\bar{b}}  \ ,
\end{equation}
where $H^{\alpha\bar{\beta}}$ is an arbitrary matrix and
$S^{a}_{\alpha}$ are the global sections of $\Vsheaf\otimes \cL^k$. As
in \eqref{sec_prod}, this fiber metric induces an inner
product on the space of sections, $S^{a}_{\alpha}$, given by
\begin{equation}
  <S_{\beta}|S_{\alpha}>=\frac{N_k}{\text{Vol}_{CY}}
  \int_{X}
  S_{\alpha}^{a}(G^{a\bar{b}})^{-1}\bar{S}^{\bar{b}}_{\bar{\beta}}\dVol_{CY}
  =\frac{N_k}{\text{Vol}_{CY}}\int_{X} 
  S_{\alpha}^{a}
  (S^{a}_{\gamma}H^{\gamma\bar{\delta}}{\bar
    S}^{\bar{b}}_{\bar{\delta}})^{-1}
  \bar{S}^{\bar{b}}_{\bar{\beta}}\dVol_{CY}
\  .
\label{coffee2}
\end{equation}

With this definition of the inner product on sections, one can give
a natural generalization of the T-operator \eqref{t_oper}. This
generalization,
\begin{equation}\label{t_gen}
  T(H)_{\alpha\bar\beta} =\frac{N_k}{\text{Vol}_{CY}}
  \int_X       
  S_\alpha 
  \Big( S^\dagger H S \Big)^{-1}
  \bar S_{\bar \beta}
  \dVol_{CY} \ ,
\end{equation}
was introduced in~\cite{MR2154820} and studied
numerically in~\cite{Douglas:2006hz}.
Note that when $\Vsheaf$ is a line bundle, \eqref{t_gen}
reduces to \eqref{t_oper} and we return to the case of a balanced
embedding into $\mathbb{P}^{N_{k}-1}$. As in the previous section, we
will describe how the iteration of the generalized T-operator can
produce a fixed point which describes an Hermitian-Einstein bundle
metric.

In~\cite{MR2154820}, it was shown that the bundle $\Vsheaf$ is
Gieseker stable\footnote{Let $\cL$ be an ample line bundle. For any
  torsion-free sheaf $\cF$ define the Hilbert polynomial with respect
  to $\cL$ as
  \begin{equation}\label{gieseker}
    p_{\cL}(\cF)(n)=\frac{\chi(\cF\otimes \cL^{n})}{rank(\cF)}
  \end{equation} 
 where $\chi(\cF\otimes \cL^{n})$ is the index of $\cF\otimes \cL^{n}$. Given two polynomials $f$ and $g$, we will write $f \prec g$ if
  $f(n) < g(n)$ for all $n \gg 0$. Then a bundle $\Vsheaf$ is said to
  be Gieseker stable if for every non-zero torsion free subsheaf $\cF
  \subset \Vsheaf$, $p_{\cL}(\cF) \prec
  p_{\cL}(\Vsheaf)$~\cite{huybrechts}.} if and only if the $k$-th
embedding (defined by $\Vsheaf\otimes \cL^{k}$ as in \eqref{V_embed})
can be moved to a ``balanced'' place. That is, if there exists an
orthonormal section-wise metric on the twisted bundle such that
\begin{equation}
  \label{T_gen_bal}
  (T(H)_{\alpha\bar\beta})^{-1} 
  =
  H^{\alpha\bar{\beta}}
\end{equation}
is a fixed point of the generalized T-operator 
We can use this
special metric on $\Vsheaf \otimes \cL^k$ to define an Hermitian metric
on $\Vsheaf$ itself. Let $G_{\cL}$ denote the Hermitian metric on
$\cL$ and $G^{(k)}$ the balanced metric on $\Vsheaf \otimes
\cL^{k}$. Then
\begin{equation}
  \label{tensor_met}
  G_{k}=G^{(k)}\otimes G_{\cL}^{-k}
\end{equation}
is an Hermitian metric on $\Vsheaf$. This appears in the following
important theorem~\cite{MR2154820}.
\begin{theorem}[Wang]
  \label{wang1}
  Suppose $\Vsheaf$ is a rank $n$, Gieseker stable bundle. If $G_k \to
  G_{\infty}$ as $k \to \infty$, then the metric $G_\infty$ solves the
  ``weak Hermite-Einstein equation''
  \begin{equation}
    \label{weak_hym}
    g^{i \bar j} F_{i \bar j}
    =
    \left( 
      \mu + \frac{\overline{R}-R}{2}
    \right)
    \mathbf{1}_{n\times n}
  \end{equation}
  where
  \begin{itemize}
  \item $R$ is the scalar curvature.
  \item $\overline{R} = \int R \sqrt{\det g} \;
    \text{d}^{2d}\!x$. This vanishes (for any K\"ahler metric) on a
    manifold with vanishing first Chern class.
  \end{itemize}
\end{theorem}

Procedurally, the process of obtaining the Hermitian-Einstein fiber metric on a slope-stable bundle, $\Vsheaf$, is very similar to that outlined for the Ricci-flat connection in Section \ref{sec:metric} - for each value $k$ of the twisting, we iterate the T-operator associated with the embedding defined by $H^0(X,\Vsheaf \times \cL^{k})$ until a fixed point is reached. Then, by Theorem \ref{wang1}, in the limit that $k \to \infty$ the induced connection solves \eref{weak_hym}. However, there is an immediate and important difference between this generalized algorithm and Donaldson's algorithm for Ricci-flat metrics. Note that while all slope-stable bundles are Gieseker
stable~\cite{huybrechts}, the converse does not hold--not all Gieseker
bundles are slope-stable. That is, there certainly exist cases
where the iteration of the T-operater {\it does not}
converge (at a given $k$). However, if $\Vsheaf$ is a slope-stable holomorphic bundle,
then the iteration $H_{n+1} = T(H_n)^{-1}$ {\it will} converge at each $k$, and
in the limit that $k \to \infty$, produce the Hermitian bundle metric $G_{\infty}$ satisfying
\eqref{weak_hym} (via its associated field strength defined in
\eqref{Amath} and \eqref{herm_met}). Moreover, in the case where the
Calabi-Yau metric $g^{i \bar j}$ is Ricci-flat, \eqref{weak_hym}
simply reduces to \eqref{hym_genr}. Thus, we have found a solution
to the standard Hermitian-Einstein equations.

\subsection{Untwisting}
\label{sec:untwist}

Despite having described a balanced embedding associated with the twisted bundle
$\Vsheaf\otimes \cL^k$, and an Hermitian metric $G_{\infty}$ satisfying
\eqref{weak_hym}, our task is not yet complete. Thus far, the
discussion has been completely general and could be applied to any
stable $U(n)$ bundle. However, for the physics associated with an $E_8
\times E_8$ heterotic string theory, we are ultimately interested in
the explicit connection, $A$, on the $SU(n)$ bundle $\Vsheaf$ itself
satisfying the slope-zero Hermitian Yang-Mills equations,
\eqref{the_hym}. Since the process of twisting $\Vsheaf$ by a line
bundle $\cL^{k}$ in the above construction clearly modifies the
trace-part of the connection, one must subtract this line bundle
contribution to get the connection on $\Vsheaf$ \emph{only}. To do this,
we have to separately compute the balanced metric, $G_{\cL}$, on
$\Lsheaf$.

Each of the bundle metrics, $G_{\cL}$  on $\cL$
and $G^{(k)}$ on $\Vsheaf\otimes \cL^{k}$, as well as the Calabi-Yau metric
$g^{i\bar{j}}$ must be approximated numerically.  The computation of
the Hermitian Yang-Mills connection in \eqref{the_hym} rests on three
finite-dimensional approximations:
\begin{enumerate}
\item The degree $k_g$ at which we compute the metric,
\item The degree $k_H$ used to compute the twisted connection
\item The degree $k_h$ used to compute the $U(1)$ part of the
  connection that must be subtracted to obtain the final
  $SU(n)$ connection on $\Vsheaf$.
\end{enumerate}
Since the Hermitian Yang-Mills connections are unique (for a given
choice of Calabi-Yau metric and geometric moduli), if we can
approximate the connections on $\Vsheaf \otimes \cL^{k_{H}}$ and
$\cL^{k_{h}}$ with sufficient accuracy, then we can compute them
numerically in a completely independent way. In terms of the metrics
on $\Vsheaf \otimes \cL^{k_{H}}$ and $\cL^{k_{h}}$, we must find
\begin{itemize}
\item  the metric on $\Vsheaf \otimes
  \Lsheaf^{k_H}$, $G^{(k_H)}=(S^\dagger H S)$, where $k_{H}$ denotes the finite number of
  iterations that will be performed to compute the connection.
\item  the metric on $\Lsheaf^{k_h}$, $(G_{\cL})^{k_h}=(s^\dagger h s)$
  (and hence $G_{\cL}=(s^\dagger h s)^{1/k_{h}}$), where $k_h$ will
  denote the number of iterations to approximate the connection.
\end{itemize}
Then, as in \eqref{tensor_met} we find that
\begin{equation}
  G
  =
  G^{(k_H)}\times G_{\cL}^{-k_H}
  =
  \Big( S^\dagger H S\Big) \Big( s^\dagger h s\Big)^{-k_H/k_h}
\end{equation}
is the fiber metric \eqref{Gdef} on $\Vsheaf$. As before, $S\in
H^0(X,\Vsheaf \otimes \cL^{k_H})$ and $s \in H^0(X, \cL^{k_h})$.

Using \eqref{Amath} and \eqref{herm_met}, in terms of the Hermitian
metric the connection on $\Vsheaf$ is then simply
\begin{equation}\label{Auntwist}
  \begin{split}
    A(\Vsheaf) 
    =&\;
    \partial
    \left[ 
      \Big( S^\dagger H S\Big) 
      \Big( s^\dagger h s\Big)^{-k_H/k_h}
    \right]
    \Big( S^\dagger H S\Big)^{-1}
    \Big( s^\dagger h s\Big)^{k_H/k_h}
    \\=&\;
    \Big( \partial (S^\dagger H S) \Big) 
    \Big( S^\dagger H S\Big)^{-1}
    - \frac{k_H}{k_h}
    \Big( \partial( s^\dagger h s) \Big)
    \Big( s^\dagger h s\Big)^{-1}
    \\=&\;
    A(\Vsheaf \otimes \Lsheaf^{k_H})
    - \frac{k_H}{k_h}
    A(\Lsheaf^{k_h}) \ .
  \end{split}
\end{equation}
That is, one can ``untwist'' the connection simply by subtracting the
trace of the $U(n)$ connection to produce an $SU(n)$ connection. The
degrees of approximation $k_{H}$ and $k_{h}$ can, in principle, be
chosen independently, as long as both are sufficiently large. We will
discuss the choice of the line bundle $\cL$ and twisting degree $k_h$
in more detail in the next subsection.

The curvature is given by
\begin{equation}
  \label{eq:Funtwisted}
  F^{(0,2)} = F^{(2,0)} = 0
  ,\quad
 g^{i \bar j} F_{i \bar j}  = g^{i \bar j} \partial_{\bar j} A_i
  =
  g^{i \bar j} \partial_{\bar j} \partial_i \ln 
  \Big( S^\dagger H S\Big) \Big( s^\dagger h s\Big)^{-k_H/k_h} .
\end{equation}    

In summary, the following is an outline of the computation the
Hermitian Yang-Mills connection (that is, the solution to
\eqref{the_hym}).
\begin{enumerate}
\item Following Donaldson's algorithm derived in
  \autoref{sec:donaldson}, approximate the unique Ricci-flat metric
  $g^{i\bar{j}}$ on $X$ to the desired degree (determined by $k_g$,
  the iteration parameter of Theorem 2, \eqref{g_approx}).
\item For a given holomorphic vector bundle $\Vsheaf$, choose an
  ample line bundle $\mathcal{L}$ on $X$ and a degree $k_{H}$ (that
  is, a twisting of the vector bundle $\Vsheaf \otimes
  \mathcal{L}^{k_{H}}$) at which to compute the ``balanced'' fiber
  metric \eqref{T_gen_bal}
\item Calculate a basis $\{S_{\alpha}\}_{\alpha=0}^{N_{k_{H}}-1}$ for
  $H^0(X,\Vsheaf \otimes \cL^{k_{H}})$ at the chosen $k_{H}$.
\item Choose an initial, invertible, Hermitian matrix,
  $H^{\gamma\bar{\delta}}$ for the ansatz \eqref{G_anzatz}. Perform
  the numerical integration to compute the T-operator in
  \eqref{t_gen}.
\item Set the new $H^{\alpha\bar{\beta}}$ to be
  $H^{\alpha\bar{\beta}}=(T_{\alpha\bar{\beta}})^{-1}$.
\item Return to item $3$ and repeat until $H^{\alpha\bar{\beta}}$
  approaches its fixed point. In practice, this convergence occurs in
  less than $10$ iterations and does not depend on the initial choice
  of $H^{\alpha\bar{\beta}}$. At this point, we have obtained the
  Hermitian metric on $\Vsheaf \otimes \cL^{k_{H}}$ to the desired
  accuracy.
\item Approximate the Hermitian metric on $\cL$ to a chosen degree
  $k_{h}$ by repeating steps 2 to 6 for $\cL^{k_{h}}$
\item Compute the ``untwisted'' connection and field strength via
  \eqref{Auntwist} and \eqref{eq:Funtwisted}
\end{enumerate}
We turn now to a discussion of the accuracy of these numerical
approximations.

\subsection{Measuring the Error}
\label{sec:error}

As discussed above, in a computer implementation of the generalized
Donaldson's algorithm one must rely on three expansion parameters. Two
are associated with the approximation of the balanced Hermitian metrics on
$\Vsheaf\otimes \cL^{k_{H}}$ and $\cL^{k_{h}}$, and the third with the Calabi-Yau
metric calculated with parameter $k_g$. Having implemented this, we
would like to know how far our numerical connection for fixed  $k_g$, $k_H$, $k_h$ deviates from the
exact Hermitian Yang-Mills connection \eqref{the_hym}.

$F^{0,2}$ and $F^{2,0}$ are automatically zero with our ansatz
\eqref{Amath} for the connection. Hence, for an $SU(n)$ bundle we have
only to test how far the ``color matrix'' $g^{i\bar j}F_{i\bar j} $
deviates from the zero matrix. Note that the matrix entries depend on
the chosen frame. Thus the appropriate invariant quantity to consider
for such an Hermitian matrix is its (real) eigenvalues
\begin{equation}\label{lambdas}
  g^{i \bar j}F_{i \bar j} 
  ~\sim~
  \begin{pmatrix}
    \lambda_1 \\ & \lambda_2 \\ & & \ddots \\ & & & \lambda_n 
  \end{pmatrix} \ .
\end{equation}
For an $SU(n)$ bundle, we expect
\begin{equation}
  \lambda_i \to 0~~\forall~~ i
\end{equation}
if and only if the connection is approaching a solution to
\eqref{the_hym}. We begin by investigating the eigenvalues of the
``color matrix'' at a point. An integrated error
measure will be discussed in the next subsection. To illustrate these concepts, we turn 
to a simple example of a stable bundle.

\subsection{An Example}
\label{eg_sec}

To begin, consider the Quartic $K3$ of
\autoref{sec:QuarticK3}. We define an explicit such hypersurface in
$\mathbb{P}^3$ by 
\begin{equation}\label{spec_k3}
  \Big\{
  z_0^4+z_1^4+z_2^4+z_3^4-2 z_0 z_1 z_2 z_3
  =0
  \Big\}
  ~\subset
  \CP^3_{[z_0:z_1:z_2:z_3]} \ ,
\end{equation}
where $z_i$ denote the coordinates of $\mathbb{P}^3$. The Ricci-flat
metric on this manifold was computed up to degree $k_g = 10$ in
\autoref{sec:QuarticK3} and the results plotted in Figures
\ref{fig:MetricError-K3-sigma}, \ref{fig:MetricError-K3-curvature} and \ref{fig:MetricError-K3-EH}.

On this Calabi-Yau twofold, we define the following rank $2$,
holomorphic vector bundle with structure group $SU(2)$. This sample
bundle is defined through the so-called ``monad''
construction~\cite{okonek, Anderson:2007nc, Anderson:2008uw,
  Anderson:2009mh}
\begin{equation}
  \label{k3_eg1}
  0 
  \longrightarrow
  \cO(-3) 
  \stackrel{f}{\longrightarrow} 
  \cO(-1)\oplus \cO(-1) 
  \oplus \cO(-1) 
  \longrightarrow 
  \Vsheaf_{\text{stable}} 
  \longrightarrow
  0 \ .
\end{equation}
Here $\Vsheaf_{\text{stable}}$ is defined as the cokernel of the map $f=(x_{0}^{2},
x_{1}^{2}, x_{2}^{2})$ between the direct sums of line
bundles. 
Using the techniques of~\cite{huybrechts,
  Anderson:2008ex}, it is straightforward to prove that $\Vsheaf_{\text{stable}}$ in
\eqref{k3_eg1} is slope-stable. Hence, by \autoref{wang1}, we
expect the T-operator to converge.

As discussed in \autoref{sec:untwist}, to apply the
generalized Donaldson algorithm one must define the embedding $i_k: X
\to G(n,N_{k_{H}})$. To do this, we must compute the global sections
of the twisted line bundle $\Vsheaf \otimes \cL^{k_H}$ for some ample
line bundle $\cL$. Fortunately, for the bundle defined in
\eqref{k3_eg1} the global sections $H^0(X,\Vsheaf \otimes \cL^{k_H})$
can easily be computed as follows. Letting $\cL^{k_H}=\cO(m)$, from
\eqref{k3_eg1} we induce the twisted short exact sequence
\begin{equation}
  0 
  \longrightarrow
  \cO(m-3) 
  \stackrel{f}{\longrightarrow}
  \cO(m-1)^{\oplus 3} 
  \longrightarrow  
  \Vsheaf_{\text{stable}}\otimes \cO(m) 
  \longrightarrow
  0 \ .
\end{equation}
Then the global sections are given simply as the cokernel
\begin{equation}\label{h0_def}
  H^0(X,\Vsheaf \otimes \cO(m))
  = 
  \frac{
    H^0(X,\cO(m-1)^{\oplus 3})
  }{
    f(H^0(X,\Vsheaf \otimes \cO(m-3)))
  } \ ,
\end{equation}
where both parts of this quotient are the global sections of
sums of ample line bundles when $m\geq 3$. Furthermore, on the Quartic
$K3$ the global sections of the line bundle $\cO(n)$ can be 
computed as a simple polynomial space of dimension
\begin{equation}\label{coh_on_k3}
  h^0(X,\cO(n))=
  \displaystyle
  \begin{cases}
    0&  n<0 \\
    \binom{n+3}{3}&  0\leq n < 4 \\
    \binom{n+3}{3}-\binom{n-1}{3} & n\geq 4
  \end{cases}
\end{equation}
With these definitions in hand, one can compute a basis of
polynomials of $H^0(X, \Vsheaf \otimes \cL^{k_{H}})$ of the degree given
by \eqref{coh_on_k3}.

Following the algorithm developed in the preceeding sections, we first
compute $g^{i\bar j}F_{i\bar j}$ at the randomly chosen point
\begin{equation}\label{k3_point}
  P = 
  \big[1: 0.707124+0.707124 i :0.1:0\big] \ .
\end{equation}
The bundle $\Vsheaf$ is rank $2$, so there are two eigenvalues
$\lambda_{1},\lambda_{2}$ in \eqref{lambdas}. Define the following
extremal quantities.
\begin{equation}
  \begin{split}
    |\lambda|_\text{max} &= \max\big\{ |\lambda_1|, |\lambda_2| \big\} 
    \\
    |\lambda|_\text{min} &= \min\big\{ |\lambda_1|, |\lambda_2| \big\}  \ .
  \end{split}
\end{equation}

\begin{figure}[htbp]
  \begin{center}
    \input{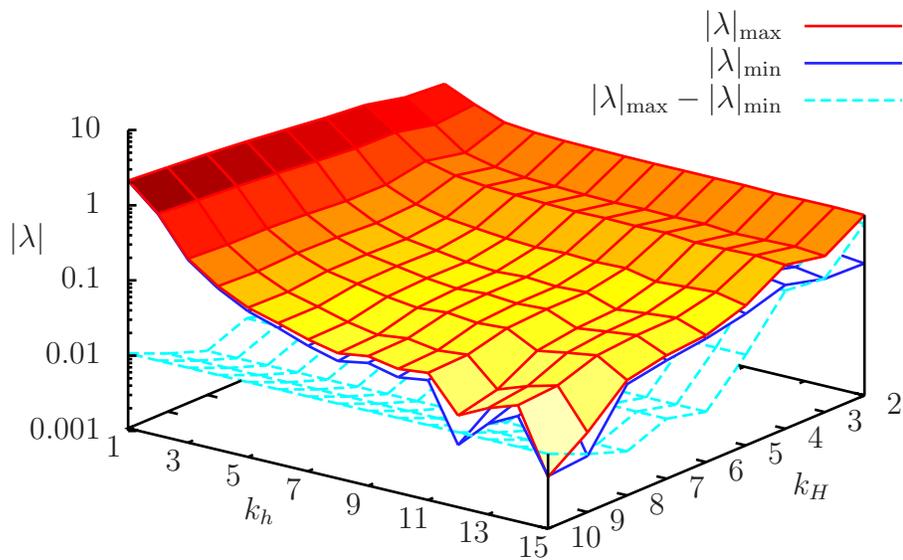}
  \end{center}
  \caption{The eigenvalues of $g^{i \bar j}F_{i\bar j}(P)$. The
    actual solution to the Hermitian Yang-Mills connection is
    characterized by having $\lambda=0$ at each point. We see that the
    numerical solution approximates this in the limit as $k_H$, $k_h
    \to \infty$.}
  \label{fig:ErrorAtP}
\end{figure}

As we iterate the twisting degrees $k_{H}$ and $k_{h}$ associated with the
connections on $\Vsheaf \otimes \cL^{k_{H}}$ and $\cL^{k_{h}}$
respectively, we can plot the eigenvalues of \eqref{lambdas} at the
point \eqref{k3_point}.  For this example, the connection T-operator
was iterated $7$ times, numerically integrating over \comma{1000000}
points.
As expected, we find that $\lambda_i \to 0$ as $k_{h},k_{H} \to
\infty$. The results are shown in \autoref{fig:ErrorAtP}. 
Inspecting this graph, we note several important features. First, 
observe that the eigenvalues will converge to zero along {\it any}
diagonal ray, that is, in any limit as $k_h, k_H \to \infty$.  Second,
note that there is clearly some variation in the rate at which the
eigenvalues along these rays converge to zero. We will define the
optimal choice in the next subsection.

\subsection{Subtracting the Trace}
\label{sec:det}

As we saw in \autoref{sec:untwist}, after twisting up the vector
bundle to $\Vsheaf \otimes \Lsheaf^{k_H}$ there is a choice in how we
compute the connection on the line bundle $\cL$. As long as there is
some limit (for example, $k_h\to\infty$ and picking the balanced
connection) where the connection on the line bundle approaches
constant curvature, we will eventually approximate the slope-zero
Hermitian Yang-Mills connection on $\Vsheaf$. Of course, some choices
of line bundles work better than others. In this subsection, we
discuss the optimal choice--the determinant line bundle of
$\Vsheaf\otimes\Lsheaf^{k_H}$ with the induced connection. The
determinant line bundle is defined as $\wedge^{n}
(\Vsheaf\otimes\Lsheaf^{k_H})$. Taking this line bundle as the ``untwisting" line bundle and choosing $k_h = \rank(\Vsheaf) k_H$, we then
choose the Hermitian metric
\begin{equation}
  ( G_{\cL})^{k_h} 
  =
  \det(G^{(k_H)})
  =
  \det \big( S^\dagger H  S\big)
\end{equation}
on $(\wedge^{n}
(\Vsheaf\otimes\Lsheaf^{k_H}))^{k_h}$. 

\begin{figure}[htbp]
  \begin{center}
    \input{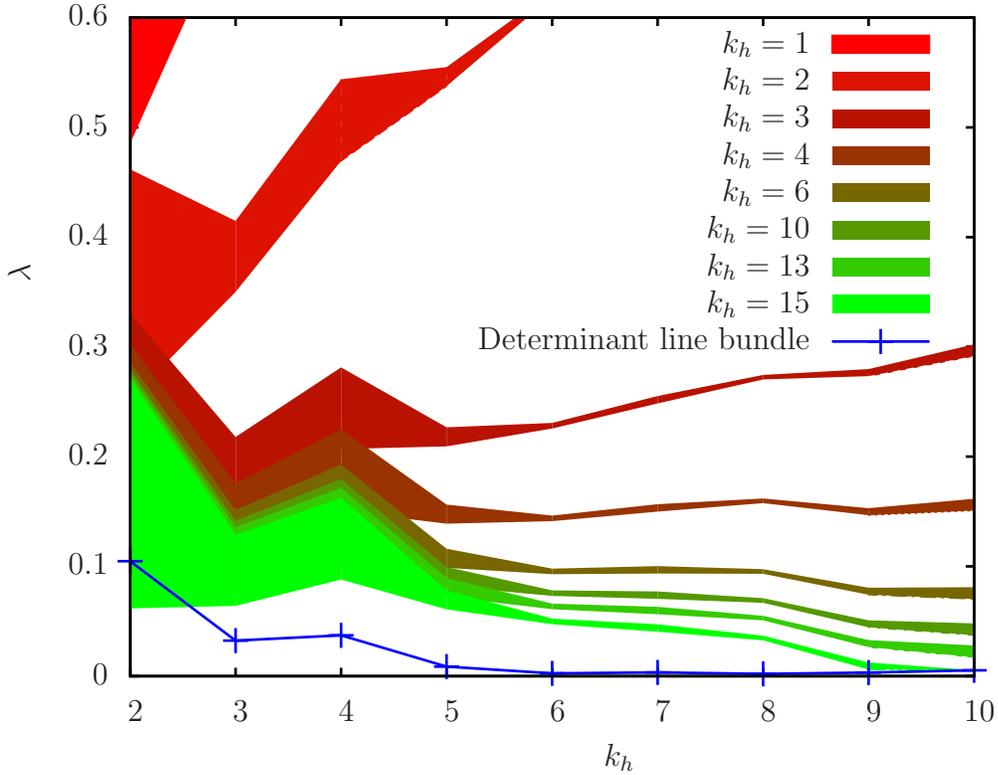}
  \end{center}
  \caption{The eigenvalues of $g^{i \bar j}F_{i\bar j}(P)$ of the rank
    $2$ bundle $\Vsheaf$ on the Quartic for different choices of
    ``untwisting line bundle'' $\cL^{k_{h}}=\cO(1)^{k_{h}}$ (same data
    as in \autoref{fig:ErrorAtP}) as well as the optimal untwisting
    with the determinant line bundle. The upper and lower boundaries
    of the colored bars are $|\lambda|_\text{max}$ and
    $|\lambda|_\text{min}$, respectively. In the determinant line
    bundle case, the two eigenvalues are $\pm \lambda$ and their
    magnitude is exactly equal at each value of $k_h$. This magnitude is shown by the blue crosses.}
  \label{fig:ErrorAtPdet}
\end{figure}

Let $\lambda^{(k_H)}$ be the ``twisted'' eigenvalues of $g^{i\bar
  j}F^{(k_H)}_{i\bar j} $ on $\Vsheaf\otimes \Lsheaf^{k_{H}}$, and let
$\lambda$ be the corresponding eigenvalues of $ g^{i\bar j}F_{i\bar
  j}$ on $\Vsheaf$ after ``untwisting''.  Then, from
\eqref{eq:Funtwisted} we have
\begin{equation}
  \lambda_i 
  = \lambda^{(k_H)}_i - 
  \frac{1}{\rank \Vsheaf} g^{i\bar j} \tr F^{(k_H)}_{i\bar j}
  = \lambda^{(k_H)}_i - \frac{\sum_j
    \lambda^{(k_H)}_j}{\rank \Vsheaf}
  .
\end{equation}
Therefore, the effect of this untwisting is precisely to subtract, at each
point, the average of the eigenvalues. In this sense, using the
determinant line bundle is the optimal way to untwist. Furthermore, with this choice of untwisting, the approximate connection on $\Vsheaf$ is $SU(n)$ at each value of $k_h$ (as opposed to the other choices for ``untwisting" line bundles for which it is only as $k_h$ becomes large that the connection approaches $SU(n)$, see Figure \ref{fig:ErrorAtPdet}).
In \autoref{fig:ErrorAtPdet}, we compare the untwisting done with
various line bundles $\cL^{k_{h}}=\cO(1)^{k_h}=\Osheaf(k_h)$ on the
Quartic with the untwisting by the determinant line bundle. The
determinant line bundle clearly produces the most rapid convergence to
zero eigenvalues. Therefore, for the remainder of this paper, regardless of the choice of ``twisting" line bundle, $\cL$, we will always use
the determinant line bundle $\wedge^n(\Vsheaf \otimes \cL^{k_H})$ to untwist.

\subsection{Integrated Error Measure}
\label{sec:tau}

To define a true error measure for the approximation to the Hermitian
Yang-Mills connection, we must test the approximation at all points
(as was done for the error measures for the Ricci-flat metric in
\autoref{sec:ricci}) and, hence, integrate over $X$. To do this, define the error
measure
\begin{equation}
  \label{eq:taudef}
  \tau(A_\Vsheaf) = 
  \frac{1}{2\pi}
  \frac{
    k_g
  }{
    \Vol_{k_{g}}
    \rank(\Vsheaf)
  }
  \int_X \Big( \sum |\lambda_i| \Big) 
  \sqrt{g} \mathop{d}\nolimits^{2d}\!x \ .
\end{equation}
This is simply a global check of the eigenvalues in
\eqref{lambdas}. To understand the normalization in \eqref{eq:taudef},
first note that $g^{i\bar j}$ scales as $\tfrac{1}{k_g}$ with the
degree $k_g$ of the metric computation. This explains the prefactor
$k_g$ in \eqref{eq:taudef}. We also divide by the metric volume
$\Vol_{k_{g}}$ to cancel the scaling of the volume element $\sqrt{g}$. Finally,
observe that
\begin{itemize}
\item If all eigenvalues are positive, then $\tau(A_\Vsheaf)=\int
  c_1(\Vsheaf) \wedge \omega^{D-1} \in \Z$. This is the reason for
  the prefactor $\tfrac{1}{2\pi}$, since $[\omega_k]=2\pi
  c_1(\cL^k)$.
\item If $c_1(\Vsheaf)=0$, then $\tau_{k_{H}}(A_\Vsheaf)\longrightarrow 0$ measures convergence to
  the Hermitian Yang-Mills connection. The eigenvalues must occur with both
  signs.
\end{itemize}
Since $c_1(\Vsheaf \otimes \cL^{k_{H}})=c_1(\Vsheaf)+nk_{H} c_1(\cL)$
is the first Chern class of $\Vsheaf \otimes \cL^{k_{H}}$ (with
$n=rank(\Vsheaf)$), we can predict the values of
$\tau^{(tw)}_{k_{H}}$ for the twisted bundle $\Vsheaf \otimes
\cL^{k_H}$ as we increase $k_{H}$. Hence, we can learn something from
evaluating the $\tau$-integral for the twisted as well as for the
untwisted connection. Define
\begin{equation}
  \begin{split}
    \tau^\text{(tw)}_{k_H} 
    =&\; 
    \tau\big( A_{\Vsheaf \otimes \Lsheaf^{k_H}} \big) 
    ,\\    
    \tau_{k_H} 
    =&\; 
    \tau\big( 
    A_{\Vsheaf \otimes \Lsheaf^{k_H}} - 
    \tfrac{1}{\rank \Vsheaf} 
    A_{\det(\Vsheaf \otimes \Lsheaf^{k_H})} 
    \mathbf{1}_{\rank \Vsheaf} \big)     
  ,
  \end{split}
\end{equation}
where $\mathbf{1}_{\rank \Vsheaf}$ is the identity matrix in the gauge indices. That is, $\tau^\text{(tw)}_{k_H}$ is the (properly normalized)
integral over the eigenvalues of the twisted-up bundle and
$\tau_{k_H}$ is the integral over the eigenvalues after untwisting
with the determinant line bundle (this expression follows immediately from \eref{Auntwist} by taking $\cL^{k_h}=(\wedge^n(\Vsheaf \otimes \cL^{k_H}))^{rank(\Vsheaf)k_H}$). Note that it implicitly depends on
the degree $k_g$ at which we computed the metric.
\begin{table}[htbp]
  \centering
  \renewcommand{\arraystretch}{1.3}
  \begin{tabular}{c|ccc}
    $k_H$ & $\int c_1(\Vsheaf\otimes \cL^{k_{H}}) \wedge \omega$ & 
    $\tau^\text{(tw)}_{k_H}$ &
    $\tau_{k_H}$ 
    \\ \hline
    $2$ &  $4$      &   $4.00057$  &    $0.904183$ \\
    $3$ &  $6$      &   $6.00329$  &    $0.765671$ \\
    $4$ &  $8$      &   $7.99855$  &    $0.532639$ \\
    $5$ &  $10$     &   $9.99944$  &    $0.331912$ \\
    $6$ &  $12$     &   $12.0012$  &    $0.229044$ \\
    $7$ &  $14$     &   $13.9983$  &    $0.170002$ \\
    $8$ &  $16$     &   $15.9989$  &    $0.132257$ 
  \end{tabular}
  \caption{The integrated error measure for the 
    stable bundle \eqref{k3_eg1} from \autoref{eg_sec}. 
    $\tau^\text{(tw)}_{k_H}$ before untwisting computes the
    integral first Chern class. The comparison between 
    the exact value for the integral Chern class and
    $\tau^\text{(tw)}_{k_H}$ is made in the first two columns. This
    comparison provides a confirmation of the validity of the
    algorithm, even before untwisting. In the final column, the error
    measure for the untwisted bundle is presented.}
  \label{tab:tautwisted}
\end{table}
In \autoref{tab:tautwisted}, we list both $\tau_{k_{H}}^\text{(tw)}$ and
$\tau_{k_{H}}$ for the rank 2 bundle defined by \eqref{k3_eg1} in \autoref{eg_sec}. In
the first column, the integral over the first Chern class of $\Vsheaf \otimes
\cL^{k_{H}}$ is listed, where $\cL=\cO(1)$ on the $K3$. As expected,
$\tau_{k_{H}}^{\text{(tw)}}$ closely matches this value. In the last column,
the untwisted error measure is presented. It clearly approaches
zero as $k_{H}$ increases.

In \autoref{fig:tauuntwisted}, we present our first explicit example of
the convergence of the generalized Donaldson algorithm for a
slope-stable bundle. We have explicitly computed the field-strength of
the $SU(2)$ bundle in \eqref{k3_eg1}. We find that the associated
integrated error measure, $\tau_{k_{H}}$ in \eqref{eq:taudef}, is converging
to zero in the limit that $k_{H} >> 1$. In addition, in
\autoref{fig:tauuntwisted} the form of the decreasing $\tau_{k_H}$
values is compared with the prediction of an $\sim 1/k_{H}$ fall-off
in the error measure predicted in~\cite{MR2154820}.


\begin{figure}[htbp]
  \begin{center}
    \input{ConnectionError-K3-curvature.tex}
  \end{center}
  \caption{$\tau_{k_H}$ for the stable bundle $\Vsheaf_{\text{stable}}$ defined in
    \eqref{k3_eg1} on the Quartic $K3$.}
  \label{fig:tauuntwisted}
\end{figure}




\section{Stable vs.\ Unstable}
\label{sec:StableVsUnstable}

In this section, we investigate the behavior of the
generalized Donaldson algorithm applied to arbitrary $SU(n)$ vector
bundles. This behavior will exhibit more complexity than we
encountered in the computation of the Ricci-flat metrics of
\autoref{sec:metric}. In applying the Donaldson algorithm to the
computation of the metric, we are guaranteed that for any compact
K\"ahler manifold $X$ with $c_1(X)=0$ the Ricci-flat metric exists.
Furthermore, the algorithm presented in \autoref{sec:donaldson} will
approximate it in the limit that $k_g \to \infty$. However, the
situation is different for vector bundles. A Hermitian-Einstein metric
does not exist for every vector bundle $\Vsheaf$ on $X$ with $c_1(\Vsheaf)=0$. As
discussed in Sections \ref{sec:Intro} and \ref{sec:generalDonaldson},
the vector bundle must be slope poly-stable for such a
Hermitian-Einstein metric to exist. Since this property is difficult to
guarantee initially, it seems prudent to ask what behavior we expect
to see if we apply this algorithm to bundles that do not, in fact, admit
an Hermitian-Einstein fiber metric? In the following, we
demonstrate that, surprisingly, the Donaldson algorithm produces
distinctive and interesting results even in the case where the bundle
is not stable. Furthermore, we will argue that in view of the difficulty
in proving slope-stability of vector bundles, \emph{the generalized
  Donaldson algorithm provides a new and attractive way of numerically
  deciding whether or not a bundle is stable}.

With this goal in mind, in the next sections we will investigate the
slope-stability properties of vector bundles and how this behavior
appears in the results of the generalized Donaldson algorithm
presented in \autoref{sec:generalDonaldson}.

\subsection{Taxonomy of Slope-Stability}
\label{sec:taxonomy}

Recall that the slope of a vector bundle $\Vsheaf$ is defined, with
respect to a given K\"ahler form $\omega$, as
\begin{equation}\label{slope2}
  \mu (\Vsheaf)
  \equiv 
  \frac{1}{\rk(\Vsheaf)}
  \int_{X}c_{1}(\Vsheaf)\wedge \omega^{d-1} 
  . 
\end{equation}
As discussed in \autoref{sec:stab}, there are four possible types
of behavior for holomorphic vector bundles. These are~\cite{huybrechts}
\begin{enumerate}
\item \textbf{Stable}: An indecomposable bundle, $\Vsheaf$, is called
  \emph{stable} if, for all subsheaves $\cF \subset \Vsheaf$,
  $\mu(\cF) < \mu(\Vsheaf)$.
\item \textbf{Poly-stable}: A bundle is called \emph{poly-stable} if
  $\Vsheaf$ is a direct sum of stable bundles with the same slope:
  $\Vsheaf=\bigoplus_i \Vsheaf_i$ with $\mu(\Vsheaf_i)=\mu(\Vsheaf)$
  $\forall~i$.
\item \textbf{Semi-stable}: A bundle is called \emph{semi-stable} if
  for all subsheaves $\cF \subset \Vsheaf$, $\mu(\cF) \leq
  \mu(\Vsheaf)$. It is \emph{properly} (or strictly) semi-stable if
  $\Vsheaf$ is semi-stable but not poly-stable (that is, it is
  indecomposable).
\item \textbf{Unstable}: An \emph{unstable} bundle is one for which
  $\mu(\cF) > \mu(\Vsheaf)$ for at least one proper subsheaf $\cF
  \subset \Vsheaf$. Unstable bundles can be either decomposable or
  indecomposable.
\end{enumerate}
Recall that a Hermitian-Einstein metric exists \emph{only} for the first
two entries in this list, that is, for poly-stable bundles (stability
being a subset of poly-stability).

Before we discuss the results of the generalized Donaldson's algorithm
in the four cases above, we must make one further observation about
slope-stability. The preceding definitions provide a measure of the
degree of substructure present in a given bundle. More precisely,
stable bundles are \emph{simple}~\cite{huybrechts}, that is,
$H^0(X,\Vsheaf \otimes \Vsheaf^{\vee})=\mathbb{C}$ and any morphism
between two stable bundles $\Vsheaf_1$,$\Vsheaf_2$ with
$rank(\Vsheaf_1)=rank(\Vsheaf_2)$ and $c_1(\Vsheaf_1)=c_1(\Vsheaf_2)$
is an isomorphism. Likewise, poly-stable bundles are direct
sums of simple objects. But properly semi-stable and unstable objects
exhibit a richer structure. Theoretically, there is a natural
framework for describing the behavior of semi-stable and unstable
vector bundles in terms of a decomposition into simple objects. We
discuss this in the next subsection.

\subsection{The Harder-Narasimhan Filtration}
\label{HS}

The following theorem makes explicit how unstable sheaves may be
described in terms of semi-stable sheaves, and semi-stable sheaves in
terms of stable sheaves~\cite{huybrechts}.

\begin{theorem}[Harder-Narasimhan]
  \label{HS_fil}
  Given a holomorphic bundle $\Vsheaf$ over a closed K\"ahler manifold
  $X$ (with K\"ahler form $\omega$), there is a filtration (called the
  Harder-Narasimhan filtration) by subsheaves
  \begin{equation}\label{hs_unstable}
    0=\cF_0 \subset \cF_1 \subset \ldots \cF_{l}=\Vsheaf
  \end{equation}  
  such that $\cF_i/\cF_{i-1}$ are semi-stable sheaves for $i=1,\ldots l$ and the slope of the quotients are ordered
  \begin{equation}
    \mu(\cF_1) > \mu(\cF_2/\cF_1)> \ldots \mu(\cF_{l}/\cF_{l-1}) \ .
  \end{equation}
  If $\Vsheaf$ is semi-stable, then there is a filtration by subsheaves (called the Jordan-H\"{o}lder filtration),
  \begin{equation}
    0=\cF_0 \subset \cF_1 \subset \ldots \cF_{l}=\tilde{\Vsheaf}
  \end{equation}  
  such that the quotients $\cF_i/\cF_{i-1}$ are all stable sheaves and have slope $\mu(\cF_{i}/\cF_{i-1})=\mu(\Vsheaf)$. In addition
  \begin{equation}\label{graded_sum}
    Gr(\Vsheaf)=\cF_1 \oplus \cF_{2}/\cF_{1}\oplus \ldots \cF_{l}/\cF_{l-1}
  \end{equation}
  is uniquely determined up to isomorphism (and is called the
  \emph{graded sum}).
\end{theorem}

One consequence of \autoref{HS_fil} is a description of the moduli
space of semi-stable sheaves. Two semi-stable bundles $\Vsheaf_1$ and
$\Vsheaf_2$ are called \emph{S-equivalent} if
$Gr(\Vsheaf_1)=Gr(\Vsheaf_2)$. The concept of S-equivalence arises when
trying to define the notion of a moduli space of sheaves. Stable
bundles correspond to unique points in their moduli space, while a
moduli space of semi-stable sheaves can only be made Hausdorff if each
point corresponds to an S-equivalence class. It is worth observing
that each S-equivalence class contains a unique poly-stable
representative (namely the graded sum \eqref{graded_sum}).

With these classifying notions in mind, we now return to our
discussion of the generalized Donaldson algorithm and the results of
the numerical scheme outlined in \autoref{sec:generalDonaldson}. If
we numerically approximate the color matrix $g^{i\bar{j}}F_{i\bar{j}}$
for the four types of bundles described in points 1)-4) above, what
will we find? What will be the eigenvalues in \eqref{lambdas}
described in \autoref{sec:error}, or the behavior of the error
measure $\tau_{k_{H}}$ defined by  \eqref{eq:taudef} in \autoref{sec:tau}?

To answer these questions, we again consider the four types of
bundles described in \autoref{sec:taxonomy} above. Depending on
the slope-stability properties of the bundle, the Harder-Narasimhan
filtration indicates that there exists a connection which will produce
a color matrix $g^{i\bar{j}}F_{i\bar{j}}$ with the following
behavior.
\begin{enumerate}
\item If $\Vsheaf$ is a poly-stable (including properly stable) bundle
  with slope $\mu$, then
  \begin{equation}\label{good_conv}
     g^{i \bar j} F_{i \bar j}\sim
    \begin{pmatrix}
      \mu & & \\ 
      &\ddots &\\
      & & \mu
    \end{pmatrix}=\mu(\Vsheaf)\bf{1}_{n\times n} \ .
  \end{equation}
  By \autoref{wang1}, in this case we expect that the generalized
  Donaldson algorithm will approximate a solution to the Hermitian
  Yang-Mills equation \eqref{gen_hym}.
\item If $\Vsheaf$ is semi-stable, then, in general, it is not a
  solution to the Hermitian Yang-Mills equations. The form of its
  color matrix will depend on the values of the bundle moduli,
  $H^1(End(\Vsheaf))$. These moduli determine where $\Vsheaf$ is
  chosen to be within it S-equivalence class. In general, the color
  matrix $g^{i\bar{j}}F_{i\bar{j}}$ will produce constant, non-equal
  eigenvalues. However, by varying the bundle moduli, $\Vsheaf$ can be
  made arbitrarily close to the unique poly-stable
  representative (the graded sum \eqref{graded_sum}) in its
  class. That is, for a properly semi-stable $SU(n)$ bundle, depending
  on the choice of bundle moduli, the eigenvalues can be made
  arbitrarily close to zero.
\item If $\Vsheaf$ is unstable, there are two possible ways in which
  the field strength fails to satisfy the Hermitian Yang-Mills
  equations. Either
  \begin{enumerate} 
  \item Its Harder-Narasimhan filtration \eqref{hs_unstable}
    consists of a sum of bundles (locally free sheaves with constant
    rank) with inequivalent slopes, $\mu_i$. In this case, the color
    matrix takes the form
    \begin{equation}\label{unstab_cons}
      g^{i \bar j} F_{i \bar j} \sim
      \begin{pmatrix}
        \mu_1 & & &\\ 
        & \mu_2 & & \\
        & & \ddots  &\\
        & & &  \mu_l
      \end{pmatrix} \ .
    \end{equation}
  \item Or its Harder-Narasimhan filtration \eqref{hs_unstable}
    contains torsion-free sheaves (whose rank can change over higher
    co-dimensional loci in the base $X$). In this case
    \begin{equation}\label{unstable_sheafy}
     g^{i \bar j} F_{i \bar j}  \sim
      \begin{pmatrix}
        \mu_1(x) & & & \\ 
        & \mu_2(x) & & \\
        &&\ddots &\\
        & & & \mu_l(x)
      \end{pmatrix} \ ,
    \end{equation}
    where $\mu_i(x)$ can vary in magnitude over the base $X$. In
    particular, at the locus in the base where the rank of $\Vsheaf$
    jumps, $F_{i\bar{j}}$ can diverge to produce a curvature
    singularity.
  \end{enumerate}
\end{enumerate}
How will this behavior appear in the generalized Donaldson
algorithm? As stated above, we are guaranteed that the algorithm
produces the physical connection only in the case when the vector
bundle is stable and the T-operator converges. But we can now ask,
what happens when we apply the algorithm to properly semi-stable or
unstable bundles? We will explore this experimentally first by
looking at sample $SU(n)$ bundles defined on the familiar Quartic
$K3$.

\subsection{Examples of Semi-Stable and Unstable Bundles on the Quartic
  $K3$}
\label{K3_egs}

We have previously seen in \autoref{eg_sec} the behavior of a
slope-stable bundle under the generalized algorithm. For the stable bundle defined in
\eqref{k3_eg1}, the algorithm produces 
convergence to an Hermitian Yang-Mills solution of the form
\eqref{good_conv}. The integrated error measure $\tau_{k_{H}}$
was shown to converge to zero in \autoref{fig:tauuntwisted}. We now
perform a simple experiment to see what happens when we apply the
generalized Donaldson algorithm in a case where \autoref{wang1} does
not apply, that is, for a non-stable bundle.

\subsubsection{A Semi-Stable Bundle}

First, let us explore the behavior of a properly semi-stable $SU(n)$
bundle. Such a bundle is not a solution to the Hermitian Yang-Mills
equations, but can come arbitrarily close to a solution as we vary the
bundle moduli. To illustrate this, consider a rank $3$ bundle over the same Quartic $K3$, \eref{spec_k3}.
We choose the $SU(3)$ monad bundle
\begin{equation}\label{k3_semi}
  \xymatrix{
    0 \ar[r] & 
    \Osheaf(-4) \ar[rr]^-{
      f_\text{semi-stable}
    } &&
    \cO(-2)\oplus \cO(-1)^{\oplus 2} \oplus \cO
    \ar[r] &
    \Vsheaf_\text{semi-stable} \ar[r] &
    0 \ ,
  }
\end{equation}
where
\begin{equation}\label{f_map_semi}
  f_\text{semi-stable} = 
  \left(
    \begin{smallmatrix}
      3uz +2xz +7yz
      \\
      \\
      \parbox{10cm}{
        \begin{math}
          \scriptscriptstyle
          ( 6u^2x +20u^2y +15uxy +5x^2y +3uy^2 +10xy^2 +18u^2z +21uxz
        \end{math}
        \flushright\vspace{-6mm}
        \begin{math}
          \scriptscriptstyle
          +7x^2z +6uyz +7xyz +7y^2z +8uz^2 +16xz^2 +10z^3 )
        \end{math}
      }
      \\[2mm]
      7u^2x +10uxy +4y^3 +6uxz +5uyz +7uz^2 +2xz^2 +10yz^2 +z^3
      \\
      8u^4 +6u^2yz +5uy^2z +5uyz^2 +19xyz^2 +7uz^3
    \end{smallmatrix}
  \right)
\end{equation}
and $x,y,z,u$ are the homogeneous coordinates on $\mathbb{P}^3$. Here
the map, $f_\text{semi-stable}$ is an element of $H^0(X,Hom(A,B))$,
where $A= \cO(-4)$ and $B=\cO\oplus \cO(-2)\oplus\cO(-1)^{\oplus 2}$.
The global sections of $\Vsheaf_{\text{semi-stable}}$ can be computed as in \eqref{h0_def} and
\eqref{coh_on_k3}. 

The Harder-Narasimhan filtration of this bundle takes the simple form
\begin{equation}
  \cO \oplus \cF \ ,
\end{equation}
where $\cF$ is a rank $2$ bundle with $c_1(\cF)=0$ defined by
\begin{equation}
  0 \to \cO(-4) \to \cO(-2)\oplus \cO(-1)^{\oplus 2} \to \cF \to 0 \ .
\end{equation}
This is an $SU(2)$ bundle, but it cannot solve the slope-zero
Hermitian Yang-Mills equations. Instead, as described in point $2$ above,
for a generic value of the bundle moduli \eqref{f_map_semi} we
expect $\tau_{k_{H}}$ to converge to a constant, finite, positive number.

\subsubsection{An Unstable Bundle with a Filtration by Line Bundles}

As our next case, consider an unstable vector bundle with a
Harder-Narasimhan filtration that is a sum of line
bundles. We select a simple example for which
$\Vsheaf_{\text{unstable}}$ itself is the following sum of line bundles over the $K3$ in \eref{spec_k3},
\begin{equation}
  \label{unstable_sum}
  \Vsheaf_{unstable~sum}=\cO(1)\oplus\cO(-1) \ .
\end{equation}
Here $\cV$ is manifestly slope-unstable since $\mu(\cO(1))>
\mu(\Vsheaf_{unstable~sum})$.  From point $3(a)$ above, we expect
the color matrix to take the form
\begin{equation}
  g^{i\bar{j}}F_{i\bar{j}} \sim \begin{pmatrix}
    1 &  \\ 
    & -1
  \end{pmatrix} \ .
\end{equation} 
This form is the same globally, since the bundle is
a direct sum. As a result, because error measure $\tau_{k_{H}}$ in
\eqref{eq:taudef} contains the sum of the absolute values of the
eigenvalues of the color matrix, we expect that as
$k_{H} \to \infty$ error measure  $\tau_{k_{H}} \sim |\lambda_1| + |\lambda_2| \to 2$.

\subsubsection{An Unstable Bundle with Sheaf Filtration}

Next, consider the case of an unstable bundle with a more
complicated Harder-Narasimhan filtration. On the Quartic $K3$ given by
\eqref{spec_k3}, we define the unstable $SU(2)$ monad bundle
$\Vsheaf_\text{unstable}$ by
\begin{equation}
  \label{unstable_k3}
  0 
  \longrightarrow
  \Osheaf(-3) 
  \stackrel{f}{\longrightarrow} 
  \cO(1)\oplus \cO(-2)^{\oplus 2}
  \longrightarrow
  \Vsheaf_\text{unstable}
  \longrightarrow
  0 \ ,
\end{equation}
where
\beq
f=(x_0^3x_2+5x_1x_2^{3}+10x_0^{2}x_1^{2}+7x_{2}x_1^{3}+4x_{0}x_3^{3}, x_1+15x_2+6x_3,2x_0+x_2+3x_1+12x_3)
\eeq 
A simple analysis along the lines of~\cite{Anderson:2007nc} (see, for example, Hoppe's Criterion), reveals that $\Vsheaf_\text{unstable}$ is a
slope-unstable bundle. It is de-stabilized by a rank $1$ sheaf,
$\cF$. Furthermore, the Harder-Narasimhan filtration of
$\Vsheaf_\text{unstable}$ is given simply by
\begin{equation}
  \label{hs_un_eg}
  \Osheaf_X(-1)\oplus \cF \ ,
\end{equation}
where $\cF$ is a rank one sheaf described by
\begin{equation}
  \label{F_subsheaf}
  0 
  \longrightarrow
  \cF 
  \longrightarrow
  \cO(2)^{\oplus 2} 
  \stackrel{g}{\longrightarrow} 
  \cO(3) 
  \longrightarrow
  0
\end{equation}
with $\cF=ker(g)$ and $c_1(\cF)=1$. This rank $1$ object is a sheaf
and not a line bundle because its rank can jump (from rank $1$ to rank
$2$) over a higher co-dimensional locus in the base $X$. This occurs
when the defining polynomial map, $g=(x_1+\ldots, 2x_0+\ldots)$ shares
common zeros with the defining polynomial of the $K3$ surface in
\eqref{k3}. It should be noted that $\Vsheaf_\text{unstable}$ itself
has no such singularities. It is a vector bundle, despite the presence
of subsheaves such as $\cF$ in \eqref{F_subsheaf}.
As a result, we expect the connection to produce a color matrix
of the type shown in \eqref{unstable_sheafy}; that is,
\begin{equation}
  g^{i \bar j}F_{i \bar j}  \sim
  \begin{pmatrix}
    \mu_{1}(x) &  \\ 
    &  -\mu_{1}(x)
  \end{pmatrix} \ .
\end{equation}

Listed above are possible expected forms for the color matrix of
several non-stable bundles. We now ask: what will the Donaldson
algorithm produce when applied to these examples? What behaviour do we expect
for the error measure $\tau_{k_{H}}$? How will the presence of
filtration-sheaves, such as \eqref{hs_un_eg}, manifest itself? Before we
investigate these questions, it should be noted that the
Harder-Narasimhan filtration of $\Vsheaf_\text{unstable}$ is {\it
  almost} the same as the form of the previous example in
\eqref{unstable_sum}. That is, the rank $1$ sheaf $\cF$ nearly
everywhere resembles the line bundle $\cO(1)$ except at points. As
mentioned above, had this been the case there would be a clear
prediction for the results of the integrated error measure. But what
happens for $\Vsheaf_{\text{unstable}}$ as defined in
\eqref{unstable_k3}? We turn now to an examination of the color matrix eigenvalues.


We present the numerical results for the three bundles above, as well as for the
stable bundle \eqref{k3_eg1} from \autoref{eg_sec}. The error measure
$\tau_{k_{H}}$ is plotted for these bundles in
\autoref{fig:CompareStableUnstableK3}. Despite the fact that these
bundles are not solutions to the Hermitian Yang-Mills equations, we
find that the way in which they fail to provide a solution are in
exact agreement with the mathematical structure of semi-stable and
unstable bundles. As expected from the Harder-Narasimhan filtrations,
the three cases shown in \autoref{fig:CompareStableUnstableK3} can be
distinguished by the behavior
\begin{equation}
  \tau \sim \int \sum|\lambda_i| \longrightarrow 
  \displaystyle
  \begin{cases}
    0~~~~~~~~~~~~~~~~~~~~~~~(\text{stable}) \\
    \text{const.} > 0~~~~~~~~~~~(\text{semi-stable}) \\
    \text{const.} > 0 \text{ or } \infty~~~~(\text{unstable})
  \end{cases}
\end{equation}
The error measure associated wth the semi-stable bundle
$\Vsheaf_{\text{semi-stable}}$ converges to a non-zero constant
value. Furthermore, the unstable sum of line bundles produces a constant
value $\tau_{k_H}=2$. Somewhat surprisingly, we find
that the error measure for $\Vsheaf_{\text{unstable}}$ converges to a
constant value of $\sim 2$, as if its Harder-Narasimhan filtration
had been $\cO(-1)\oplus \cO(+1)$ (that is, of the $\mu=const.$ type
described in \eqref{unstab_cons}) instead of the actual filtration in
\eqref{hs_un_eg}.
To understand the result for this last unstable bundle, one must look
in more detail at the computation of the field strength and the
behavior of the subsheaf $\cF$ in the 
algorithm approximating the connection. This will be explored in detail in the next subsection. We will give an
explicit discussion of how the T-operator fails to converge in this
case, and how the generalized Donaldson algorithm none-the-less
produces a connection that exhibits the correct physical
singularities.

\begin{figure}[htbp]
  \begin{center}
    \input{ConnectionError-K3-curvature-compare.tex}
  \end{center}
  \caption{The integrated error, $\tau_{k_H}$, for $SU(n)$ bundles on
    the Quartic $K3$, \eqref{k3}. Shown above are the results for 1) the
    stable $SU(2)$ bundle $\Vsheaf_{\text{stable}}$ defined in
    \eqref{k3_eg1}, 2) the semi-stable $SU(3)$ bundle
    $\Vsheaf_{\text{semi-stable}}$ defined in \eqref{k3_semi}, 3) the
    simple unstable sum of line bundles $\cO(-1)\oplus \cO(-1)$ and 4)
    the unstable $SU(2)$ bundle $\Vsheaf_{\text{unstable}}$ defined in
    \eqref{unstable_k3} with a Harder-Narasimhan filtration exhibiting
    sheaf singularities.}
  \label{fig:CompareStableUnstableK3}
\end{figure}

\subsection{Eigenvalues Along a Geodesic}\label{geodesic}

In this subsection, we explore in detail the behavior of the unstable
bundle $\Vsheaf_{\text{unstable}}$ on the Quartic $K3$. As discussed
above, while the rank of $\Vsheaf_{\text{unstable}}$ remains constant
everywhere on the $K3$ surface $X$, the rank of a de-stabilizing
subsheaf $\cF$ in \eqref{F_subsheaf} can increase over a higher
co-dimension locus in the base.

For the sheaf $\cF$, one can find this singular locus by determining
where the defining map $g$ in \eqref{F_subsheaf} goes to zero over
the Calabi-Yau \eqref{k3}. In this case, there are $4$ point-like
``instantons'', that is, points in the base for which $\cF$ jumps in rank. Such
instantons are described non-perturbatively in heterotic M-theory by a
dissolved $M5$-brane on the $K3$ surface~\cite{Ovrut:2000qi,
  Buchbinder:2002ji}. In homogeneous coordinates, these points\footnote{Such degeneracies can be easily found using computational algebraic geometry packages such as \cite{Gray:2006gn,Gray:2008zs,GLS03}.} are
\begin{equation}\label{instanton_pts}
  \begin{split}
    p_0 \;&= \big[
    1:
    -3.2279 - 1.6903 i:
    -0.7510 - 2.6031 i:
    1.3874 + 2.8736 i
    \big],
    \\
    p_1 \;&= \big[
    1:
    -3.2279 + 1.6903 i:
    -0.7510 + 2.6031 i:
    1.3874 - 2.8736 i
    \big],
    \\
    p_2 \;&= \big[
    1:
    -1.6951 - 0.3844 i:
    1.6095 - 0.5912 i:
    -1.2183 + 0.6535 i
    \big],
    \\
    p_3 \;&= \big[
    1:
    -1.6951 + 0.3844 i:
    1.6095 + 0.5920 i:
    -1.2183 - 0.6535 i
    \big].
  \end{split}
\end{equation}
To explore the behavior of the connection around, say, $p_0$, we chose
a geodesic $P(s)$ that passes through this point. The approximate
Calabi-Yau metric used for the geodesic was computed at $k_g=15$, and
the parameter $s$ is normalized such that it coincides with the path
length.

The geodesic is a solution of the differential equation
\begin{equation}
  \frac{\partial^2 P^\lambda(s)}{\partial s^2}
  + 
  \Gamma^\lambda_{\mu\nu} 
  \frac{\partial P^\mu(s)}{\partial s}
  \frac{\partial P^\nu(s)}{\partial s}
  =
  0
\  .
\end{equation}
On a K\"ahler manifold, parallel transport does not mix holomorphic
and anti-holomorphic coordinates. In other words, the Christoffel
symbol
\begin{equation}
  \Gamma^i_{j k}
  =
   g^{i \bar h} \partial_j g_{k  \bar h}
\end{equation}
has no mixed-index components. Therefore, the equation for the
geodesic simplifies to
\begin{equation}
  \frac{\partial^2 P^i(s)}{\partial s^2}
  + 
  \Gamma^i_{j k} 
  \frac{\partial P^j(s)}{\partial s}
  \frac{\partial P^k(s)}{\partial s}
  =
  0
\  .
\end{equation}
We normalize the initial velocity vector as
\begin{equation}
  1 = 
  g_{i\bar j} 
  \frac{\partial P^i(s)}{\partial s}
  \overline{
    \frac{\partial P^j(s)}{\partial s}
  }
  \Big|_{s=0}
  \ ,
\end{equation}
so that the geodesic parameter $s$ equals the metric path length. To numerically solve for  $P(s)$, we used Newton's method with a step size of $10^{-3}$. After each integration
step, we projected back onto the K3 hypersurface. 
Some sample points along our chosen geodesic are
\begin{equation}
  \begin{split}
    P(0) =&\;
    p_0,
    \\
    P(0.05) =&\; 
    \big[
    1.005 +0.001 i:
    -3.228 -1.690 i:
    -0.749 -2.593 i:
    1.387 +2.879 i
    \big],
    \\
    P(1) =&\; \big[
    1.640 +0.372 i:
    -3.228 -1.690 i:
    -0.426 -0.058 i:
    1.117 +3.389 i
    \big],
    \\
    P(2) =&\; \big[
    2.127 +0.968 i:
    -3.228 -1.690 i:
    -0.528 +2.744 i:
    0.608 +3.501 i
    \big].
  \end{split}
\end{equation}

\begin{figure}[htb]
  \begin{center}
    \input{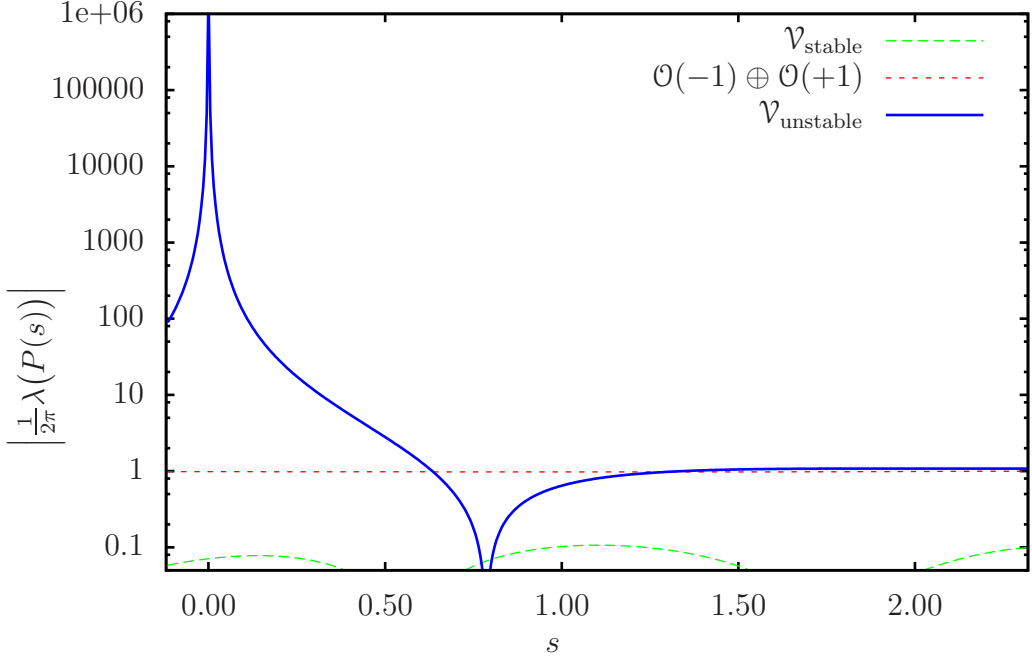}
  \end{center}
  \caption{The eigenvalues of $g^{i \bar j}F_{i \bar j}$ along the
    geodesic $P(s)$. The indecomposable, unstable bundle
    $\Vsheaf_\text{unstable}$ is the thick blue line, and one of its
    point-like instantons is located at $s=0$. For comparison, the
    eigenvalues for the balanced connection on the sum of line
    bundles \eqref{unstable_sum} are given by the red line, while that for the stable bundle \eqref      
    {k3_eg1} is shown by the
    green line.} 
  \label{fig:QuarticK3-lambda-geodesic-fit}
\end{figure}
We now compute $ g^{i \bar j}F_{i \bar j}$ for
$\Vsheaf_{\text{unstable}}$ at each point of the geodesic. Recall that
$\Vsheaf_{\text{unstable}}$ remains a bundle everywhere. We want to
know: what happens to the color matrix $g^{i\bar{j}}F_{i\bar{j}}$ near
the $\cF$-sheaf singular point $P(0)$?
Since the vector bundles we are considering here are of rank $2$,
there are two real eigenvalues. After untwisting with the determinant
line bundle, they are $\pm \lambda\big( P(s) \big)$. We plot these
eigenvalues in \autoref{fig:QuarticK3-lambda-geodesic-fit}. Note, since $\Vsheaf_{\text{unstable}}$ is an $SU(2)$ bundle, the two eigenvalues are $\lambda$ and $-\lambda$. We plot only the positive eigenvalue. For
comparison, we also compute the color matrix for the sum of line
bundles $\cO(-1)\oplus \cO(+1)$ and the stable bundle,
$\Vsheaf_{\text{stable}}$ from \eqref{k3_eg1} along the same geodesic.
Since we are interested in the accuracy of our numerical approximation
to the slope-unstable connections, we list here the details of this
preliminary calculation (in the sense that we have performed only a few iterations
of the non-converging T-operator).
\begin{itemize}
\item The Calabi-Yau metric (which determines the K\"ahler form $\omega$
  as wells as the geodesic $P$) was computed at degree $k_g=15$.
\item The metric T-operator was iterated $15$ times, numerically
  integrating over $\comma{2000000}$ points.
\item The Hermitian Yang-Mills connection $F_{i \bar j}$ was computed
  at degree $k_H = 7$.
\item The connection T-operator was iterated $30$ times, numerically
  integrating over $\comma{1000000}$ points.
\end{itemize}

By inspecting \autoref{fig:QuarticK3-lambda-geodesic-fit}, we see that the results
for the stable bundle $\Vsheaf_\text{stable}$ and the unstable sum of
line bundles $\Osheaf(-1)\oplus \Osheaf(+1)$ behave smoothly along the
geodesic. The eigenvalues for the stable bundle 
$\Vsheaf_\text{stable}$ (green dashed line) are close to zero, as they should be for a
slope-stable bundle with $c_1(\Vsheaf_\text{stable})=0$. Likewise, the
eigenvalues for the reducible bundle $\Osheaf(-1)\oplus \Osheaf(+1)$
are approximately constant and equal to the expected value
\begin{equation}
  \frac{1}{2\pi}\lambda = c_1\big( \Osheaf(1) \big) = 1
  \ .
\end{equation}
More interestingly, as predicted, the eigenvalues for the unstable
bundle $\Vsheaf_\text{unstable}$ have a pole at $P(0)$, the location
of the point-like instanton. Note that due to the log-scale in
\autoref{fig:QuarticK3-lambda-geodesic-fit}, the width of this pole appears somewhat
elongated. In reality, this divergence is very close to a
$\delta$-function, $\delta(s=0)$. Away from that point,
$\tfrac{1}{2\pi}\lambda \approx 1$. This eigenvalue was precisely the
value that was obtained in \autoref{fig:CompareStableUnstableK3} by
the general numerical integration (which did not include the points
\eqref{instanton_pts}).

As a final observation, note that the ``dip'' in the
eigenvalues, where $\lambda \to 0$ before assuming the constant
value, is a function of the fact that the bundle $\Vsheaf_{\text{unstable}}$
is not defined globally as a direct sum. That is, near the singularity
at $s=0$ the eigenvalues of the color matrix have the form
\begin{math}
 \left( \begin{smallmatrix}
    \lambda &  \\
    & -\lambda
  \end{smallmatrix}\right),
\end{math}
whereas sufficiently far from the singularity,
for $s \gtrsim 0.9$, we find that the color matrix has ``switched'' the position of
its positive and negative eigenvalues; that is,
\begin{math}
\left(  \begin{smallmatrix}
    -1 &  \\
    & 1
  \end{smallmatrix}\right).
\end{math}
In transitioning between these two configurations, the eigenvalues
must go to zero. We conjecture that the failure of the eigenvalues to
decompose globally into $+$ and $-$ values is in one-to-one
correspondence with the fact that $\Vsheaf_{\text{unstable}}$ is not defined as a direct
sum of line bundles but, rather, as an indecomposable rank 2 bundle. There is no physical significance to the point
where these eigenvalues approach zero. In fact, we will see in the
next subsections that as one increases the accuracy of our computation
(specifically, higher iterations of the T-operator and of the $k_{H}$-twisting)
the position of this zero approaches the $\delta$-function spike at
$s=0$.

\subsubsection{Iterating the T-Operator}\label{iter_geodesic}

It is clear that the T-operator \eqref{t_gen} on the matrices
$H_{\alpha\bar\beta}$ does \emph{not} converge for
$\Vsheaf_\text{unstable}$. As noted in
\autoref{sec:generalDonaldson}, the T-operator converges if and only
if the bundle is Gieseker stable and $\Vsheaf_\text{unstable}$ is
manifestly Gieseker unstable\footnote{To see that
  $\Vsheaf_{\text{unstable}}$ is not Gieseker stable, note that from
  \eqref{gieseker}, $P_{\cO(1)}(\cF)(n)=\chi(\cF\otimes
  \cO(n))=2n(2+n)$ while $P_{\cO(1)}(\Vsheaf)(n)=\chi(\Vsheaf \otimes
  \cO(n))=2(1+n^2)$. Hence, for $n \gg 1$, $P_{\cO(1)}(\Vsheaf) \prec
  P_{\cO(1)}(\cF)$ and $\Vsheaf$ is Gieseker unstable.}. Indeed, since
the T-operator has to reproduce the field-strength singularity at the
location of the point-like instanton, some components of the
$H_{\alpha\bar\beta}$ matrix must grow beyond bound as we approximate
the balanced connection.

\begin{figure}[htb]
  \begin{center}
    \input{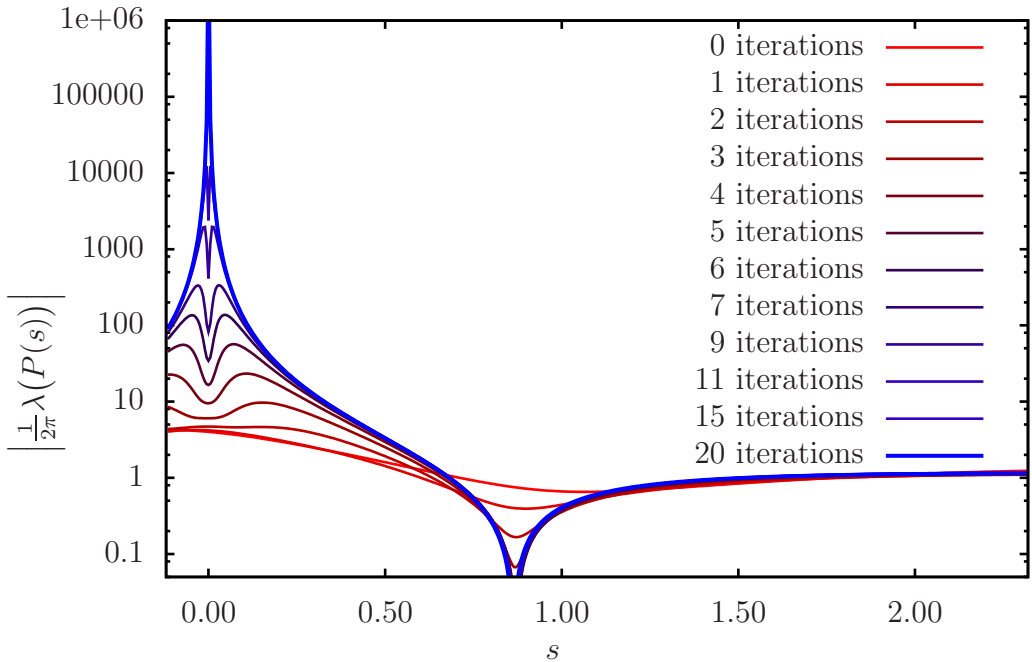}    
  \end{center}
  \caption{The eigenvalues $\lambda\big(P(s)\big)$ of $g^{i \bar j}F_{i \bar j}$
    along the geodesic $P(s)$. The point-like instanton
    is located at $s=0$. The Hermitian Yang-Mills connection was
    approximated at degree $k_H=5$, and the bundle T-operator was
    iterated between 0 and 20 times.}
  \label{fig:QuarticK3-lambda-geodesic-n_iter_G}
\end{figure}

To see this behavior, consider higher iterations of the
T-operator. We approximate the field strength $F_{i \bar j}$ at fixed
degree $k_H = 5$, and compute the T-operator by numerically
integrating using $\comma{500000}$ points. The results of the first
$20$ iterations of the T-operater are plotted in
\autoref{fig:QuarticK3-lambda-geodesic-n_iter_G}. Despite the fact that the T-operator is clearly failing to converge,
and, hence, the results of \autoref{wang1} do not apply, the numerical
evidence presented here strongly suggests that the balanced connection
is still a well-defined limit as we continue iterating the T-operator.

As a final exploration of this behavior, we perform the computation of
the connection at the limit of our accuracy, that is, for both a 
high degree of twisting and a large number of iterations of the
T-operator.

\subsubsection{Degree of Twisting}

\begin{figure}[htb]
  \begin{center}
    \input{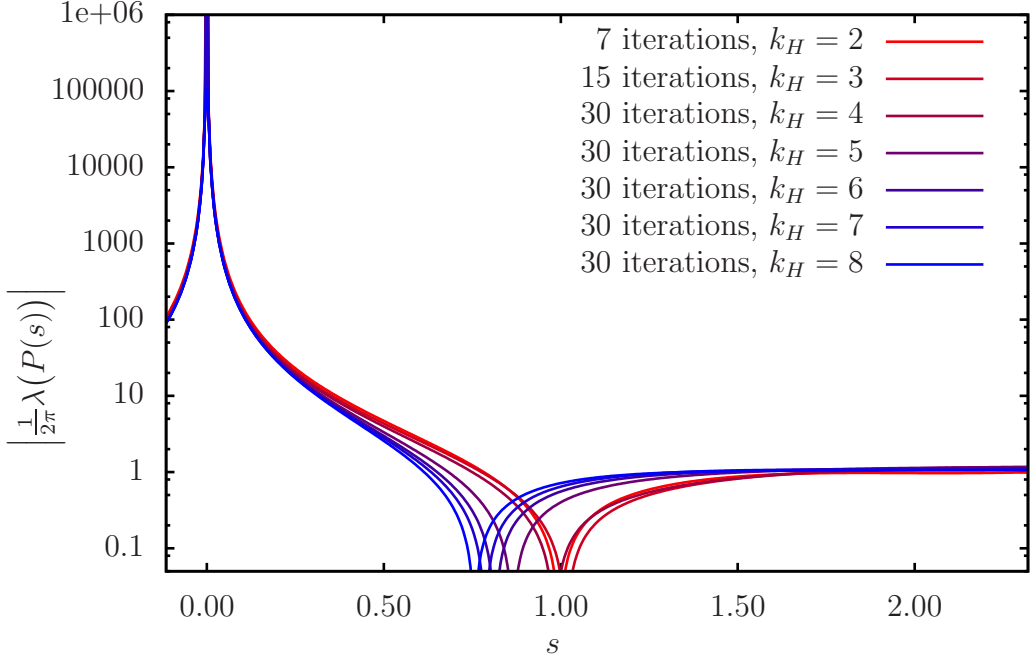}
  \end{center}
  \caption{The eigenvalues $\lambda\big(P(s)\big)$ of $g^{i \bar j}F_{i \bar j}$
     along the geodesic $P(s)$. The point-like
    instanton is located at $s=0$. The different lines correspond to
    different degrees $k_H$ at which the Hermitian Yang-Mills
    connection was approximated. The divergent peak at $s=0$ is approaching a $\delta$-funtion $\delta(s=0)$ as $k_{H} \to \infty$.}
  \label{fig:QuarticK3-lambda-geodesic-k_G}
\end{figure}

In this subsection, we compute the color matrix
$g^{i\bar{j}}F_{i\bar{j}}$ at degrees ranging from $k_H = 2$ to
$k_H=8$. The connection T-operator was numerically integrated using
$\comma{1000000}$ points. For small degrees $k_H$, some entries of
$H_{\alpha \bar\beta}$ grow quickly at each iteration, and the
iteration must be stopped before the machine precision becomes
insufficient. This is why we performed only $7$ and $15$ iterations at
$k_H=2$ and $k_H=3$. For larger degrees $k_H\geq 4$, the matrix
entries grow slowly enough that this is not an issue. The results are
shown in \autoref{fig:QuarticK3-lambda-geodesic-k_G}.

For a bundle of the form $\Vsheaf_{\text{unstable}}$ in
\eqref{unstable_k3}, we conjecture that as $k_H\to \infty$ the color
matrix can be decomposed as sum of a divergent piece (due to the sheaf
$\cF$ in \eqref{hs_un_eg} in its Harder-Narasimhan filtration) plus a
constant eigenvalue contribution (due to the ``smoothing'' of its
Harder-Narasimhan filtration to $\cO(-1)\oplus \cO(1)$ far from
the points \eqref{instanton_pts}). That is,
\begin{equation}
  (\text{point-like instanton}) + (\text{constant}~ \lambda~\text{elsewhere}).
\end{equation}
This conjecture explains the convergence of $\tau_{k_{H}}$ to a
constant value shown in \autoref{fig:CompareStableUnstableK3}. Specifically,
that
\begin{equation}
  \lim_{k_H\to\infty}\tau_{k_H}
  \Big( \Vsheaf_\text{unstable} \Big)
  = 
  \lim_{k_H\to\infty}\tau_{k_H}
  \Big( \Osheaf(-1)\oplus\Osheaf(+1) \Big)
  = 2
\end{equation}
for regions away from the point-like instantons.

In summary, the results of the past few sections have given
substantial evidence that even in the cases where a bundle $\Vsheaf$
does not solve the Hermitian Yang-Mills equations (and, hence, the
T-operator does not coverge), the generalized Donaldson algorithm
still produces a physically relevant connection. In particular, in
view of the results in  \autoref{fig:QuarticK3-lambda-geodesic-k_G} it seems
possible that even in the case of an unstable vector bundle for which
the color matrix $g^{i\bar{j}}F_{i\bar{j}}$ diverges, the generalized
Donaldson algorithm is providing an accurate approximation of the
field strength $F_{i\bar{j}}$. In the limit as $k_{H}$ increases, as
well as the iterations of the T-operator, we find that while the
T-operator is not converging, the resulting connection not only
correctly displays the field strength singularities due the sub-sheaf
singularities, but also the smooth asymptotic behavior away from any
singular loci.

With these observations in hand, we conclude that it is possible to apply
the generalized Donaldson algorithm to an arbitrary holomorphic vector
bundle and to determine its slope-stability properties by inspection
of the results, such as those shown in
\autoref{fig:CompareStableUnstableK3}. This observation provides a
useful new tool for determining the existence of supersymmetric vacua
in heterotic compactifications.

\subsection{Rank Three Bundles on the Quintic}
\label{sec:quinticHYM}

Having learned what behavior to expect from stable, semi-stable, and unstable bundles under the generalized Donaldson algorithm, we now
present examples involving higher rank bundles on a Calabi-Yau
threefold. This is the type of geometry that is ultimately of interest to us in ${\cal N}=1$ supersymmetric heterotic compactifications.

Specifically,
we consider the deformed Fermat Quintic, denoted $\mathbb{P}^4[5]$, defined by
\begin{equation}
  \label{quintic_again}
  u^5+v^5+x^5+y^5+z^5+\tfrac{5}{2} u v x y z = 0 \ ,
\end{equation}
where $u,v,x,y,z$ are the homogeneous coordinates of $\mathbb{P}^4$. As was the case on the Quartic $K3$, the K\"ahler cone of the Quintic is $1$-dimensional and all line bundles $\cL$ on $X$ can be written as $\cO(n)$ for some integer $n$. The global sections of ${\cal O}(n)$ on $X$ can be defined as a polynomial space whose dimension is given by
\begin{equation}\label{coh_on_quintic}
  h^0(X,\cO(n))=
  \displaystyle
  \begin{cases}
    0&  n<0 \\
    \binom{n+4}{4}&  0\leq n < 5 \\
    \binom{n+4}{4}-\binom{n-1}{4} & n\geq 5
  \end{cases}~~~.
\end{equation}
With these definitions in hand, one can compute a basis of
polynomials of $H^0(X, \Vsheaf \otimes \cL^{k_{H}})$ of the degree given
by \eqref{coh_on_quintic}.

As in previous sections, we will investigate three types of vector bundle that are respectively 1) stable, 2)
semi-stable and 3) unstable. We will consider two different types of unstable bundles-- one an
indecomposable rank $3$ bundle and the other an unstable sum of three
line bundles. The monad bundles are defined as follows.

\subsubsection{ A stable bundle}
The stable $SU(3)$ bundle $ \Vsheaf_\text{stable}$ is given by
  \begin{equation}
    \label{quintic_stable}
    0
    \longrightarrow
    \Osheaf(-2)^{\oplus 3}   
    \stackrel{{f_\text{stable}}}{\xrightarrow{\hspace{1cm}}} 
    \Osheaf(-1)^{\oplus 6}  
    \longrightarrow
    \Vsheaf_\text{stable} 
    \longrightarrow
    0 \ ,
  \end{equation}
  where the monad map is defined (via a map on sections) as
  \begin{equation}
    f_\text{stable} 
    =
    \left(\begin{smallmatrix}
        5 u + 24 v + 3 x + 7 y + 18 z &
        11 u + 14 v + 18 x + 14 y + 29 z &
        12 u + 16 v + 11 x + 13 y + 16 z 
        \\
        9 u + 22 v + 12 x + 14 y + 3 z &
        8 u + 4 x + 9 z &
        20 u + 33 v + 9 x + 18 y + 34 z 
        \\
        6 x + 3 y &
        3 u + v + 2 x + 13 y + 18 z &
        24 u + 23 v + 46 x + 17 y + 38 z 
        \\
        10 v + 12 y + 10 z &
        6 v + 11 x + 9 y &
        36 u + 21 v + 12 x + 37 y + 28 z 
        \\
        13 u + 26 v + 18 x + 13 y + 12 z &
        5 u + 11 v + 6 x + 9 y + 12 z &
        3 u + 18 x + 15 y + 12 z 
        \\
        20 u + 28 v + 5 x + 12 y + 17 z &
        12 u + 14 v + 28 x + 6 y + 7 z &
        10 u + 4 y + 6 z
      \end{smallmatrix}\right) \ .
  \end{equation}
  
\begin{figure}[htbp]
  \begin{center}
    \input{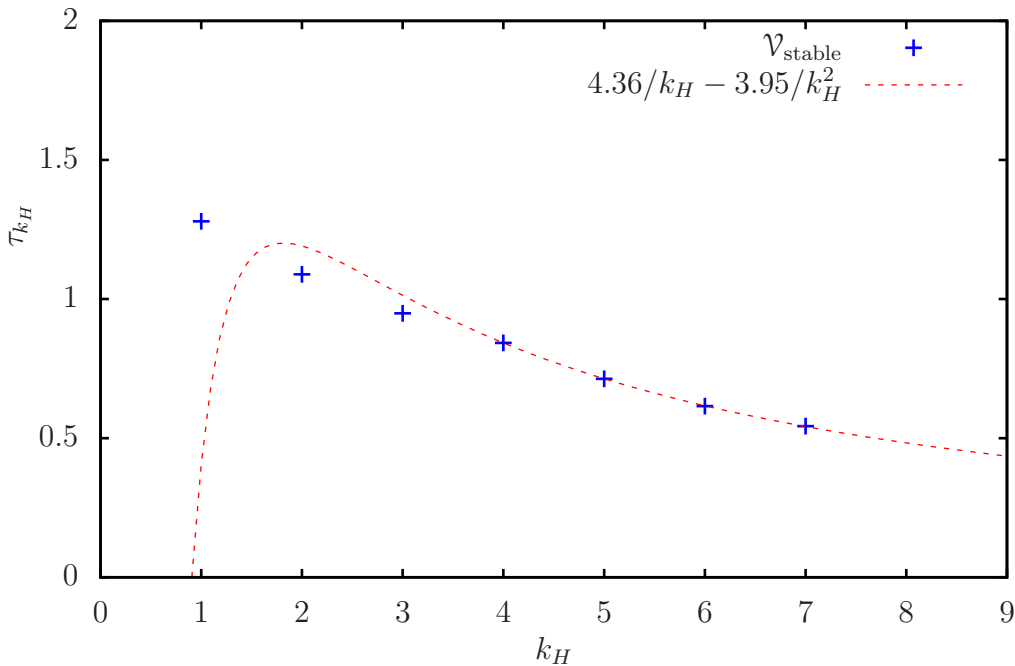}
  \end{center}
  \caption{The integrated error measure $\tau_{k_H}$ on the Quintic
    threefold for the stable bundle defined in
    \eqref{quintic_stable}.}
  \label{fig:StableQuintic}
\end{figure}

  As in the previous sections, since the K\"ahler cone of the Quintic is one-dimensional, it is straightforward to verify that $\Vsheaf_{\text{stable}}$ is slope-stable by checking that $H^0(X, \wedge^k V)=0$ for $k=1,\ldots 2$ (Hoppe's criterion \cite{huybrechts,Anderson:2008ex}). Applying the generalized Donaldson algorithm to the computation of the connection on $\Vsheaf_{\text{stable}}$, we find that the T-operator converges as expected and that the error measure, \eqref{eq:taudef}, converges smoothly to zero. The results are shown in \autoref{fig:StableQuintic}.

\subsubsection{A semi-stable bundle}
A semi-stable $SU(3)$ bundle is defined by the following short exact sequence
  \begin{equation}\label{quintic_semi}
    0 
    \longrightarrow
    \Osheaf(-3)  
    \stackrel{{f_\text{semi-stable}}}{\xrightarrow{\hspace{2cm}}}
    \Osheaf(-1)^{\oplus 3} \oplus \Osheaf 
    \longrightarrow
    \Vsheaf_\text{semi-stable}
    \longrightarrow
    0 \ ,
  \end{equation}
  where $f_{\text{semi-stable}}$ is
  \begin{equation}
    f_\text{semi-stable} 
    =
    \left(\begin{smallmatrix}
        4 u^2 + u v + 11 u x + 15 v x + 8 x^2 + 15 v y + x y + 16 u z + 
        21 v z + 26 x z + 14 y z + 36 z^2
        \\
        5 u^2 + 4 u v + 19 u x + 9 v x + 8 u y + 6 v y + 10 x y + y^2 + 
        24 u z + 21 v z + 17 x z + 9 y z
        \\
        (2 u^2 + 6 u v + 20 v v + 15 u x + 5 v x + 14 u y + 12 v y + 19 x y +
        5 v z + 4 z^2)\\
        {\scriptsize
          (4 u^3 + 10 u^2 v + 13 u v x + 15 u v y + u x y + 16 v x y +
          u y^2 + 5 x y^2 + 5 y^3 }\\
        {\scriptsize
          +6 u^2 z + 3 u v z + 6 v^2 z + 7 u x z + 
          11 v x z + 9 u y z + 8 v y z + 9 x y z + 5 y^2 z + 8 v z^2 + 
          6 x z^2 + 4 y z^2) 
        }
      \end{smallmatrix}\right) \ .
        \end{equation}
In this case, the bundle admits a subsheaf, $\cO \subset \Vsheaf_{\text{semi-stable}}$ with $\mu(\cO)=\mu(\Vsheaf_{\text{semi-stable}})$. Hence, $\Vsheaf_{\text{semi-stable}}$ is properly semi-stable, not polystable, and will not admit a solution to the Hermitian Yang-Mills equations. Its Jordan-H\"{o}lder filtration \eqref{graded_sum} is given simply by
      \begin{equation}\label{filt}
      {\cal F}\oplus \cO~,
      \end{equation}
      where the rank $2$ sheaf ${\cal F}$ is defined by
      \begin{equation}
      0 \to {\cal F} \to \cO(1)^{\oplus 3} \to \cO(3) \to 0~.
      \end{equation}
  Because the S-equivalence class containing $\Vsheaf_{\text{semi-stable}}$ contains a poly-stable representative, this semi-stable bundle may be brought arbitrarily close to a solution of the Hermitian Yang-Mills equations by varying the bundle moduli (that is, bringing $\Vsheaf_{\text{semi-stable}}$ closer to the direct sum bundle, \eref{filt}).

\subsubsection{Unstable bundles}
As in previous sections, we will compare the behavior of two unstable bundles. 

\begin{enumerate}
\item \text{\bf{An unstable direct sum:}} 

First, consider the simple unstable direct sum defined by
  \begin{equation}
  \Vsheaf_{sum}=\cO(-1)\oplus\cO(-1)\oplus\cO(2)~~.
  \label{home1}
  \end{equation}
  $ \Vsheaf_{sum}$ is a direct sum of stable objects, but because the three line bundles do not have the same slope, the sum is unstable. As in \eqref{unstab_cons}, the color matrix is given by
  \beq
  g^{i\bar{j}}F_{i\bar{j}}=\begin{pmatrix}
        -1 & &\\ 
        & -1 & \\
         & & 2
      \end{pmatrix} \ .
  \eeq
  
  The error measure, \eref{eq:taudef}, is predicted to lie at $\tau_{k_{H}}=4$.
  \item \text{\bf{An indecomposable unstable bundle:}} 
  
  We can define an unstable $SU(3)$ bundle as follows,
    \begin{equation}\label{quintic_unstable}
      0
      \longrightarrow
      \Osheaf(-2)  
      \stackrel{f_{\text{unstable}}}{\xrightarrow{\hspace{1.5cm}}} 
      \Osheaf(-1)^{\oplus 4} \oplus \Osheaf(1)
      \longrightarrow
      \Vsheaf_\text{unstable}
      \longrightarrow
      0
    \end{equation}
    with
    \begin{equation}
      f_\text{unstable} 
      =
      \left(\begin{smallmatrix}
          x+6 y+18 z
          \\
          13 u+11 x+9 y
          \\
          7 u+11 v+4 z \\
          {\scriptsize
            (4 u^3 + 10 u^2 v + 13 u v x + 15 u v y + u x y + 16 v x y +
            u y^2 + 5 x y^2 + 5 y^3 + 6 u^2 z + 3 u v z}\\
          {\scriptsize + 6 v^2 z + 7 u x z + 
            11 v x z + 9 u y z + 8 v y z + 9 x y z + 5 y^2 z + 8 v z^2 + 
            6 x z^2 + 4 y z^2)}
        \end{smallmatrix}\right) \ .
    \end{equation}
    Here $\Vsheaf_{\text{unstable}}$ is destabilized by the sub-bundle $\cO(1)$ with $\mu(\cO(1))>0$. The graded sum, \eqref{graded_sum}, associated with its Harder-Narasimhan filtration is given simply by
    \begin{equation}
    {\cal G}\oplus \cO(1)~,
    \end{equation}
    where ${\cal G}$ is a rank $3$ sheaf with $c_1({\cal G})=-1$ defined by
    \begin{equation}
    0 \to \cO(-2) \to \cO(-1)^{\oplus 3} \to {\cal G} \to 0~~.
    \end{equation}
    As discussed in Subsections \ref{K3_egs} and \ref{geodesic}, we would not expect that since $|c_1({\cal G})|=|c_1(\cO(1))|=1$, the error measure should approximate $\tau_{k_{H}}=2$.
  \end{enumerate}

In these calculations, the metric on the Quintic threefold was computed at degree
$k_g=8$, the metric T-operator was iterated $10$ times and the numerical
integration was carried out with \comma{2000000} points. In addition, the connection
T-operator was iterated $10$ times, the numerical integration used
\comma{1000000} points and the $\tau_{k_H}$ integral was computed
numerically with \comma{100000} points.

\begin{figure}[htbp]
  \begin{center}
    \input{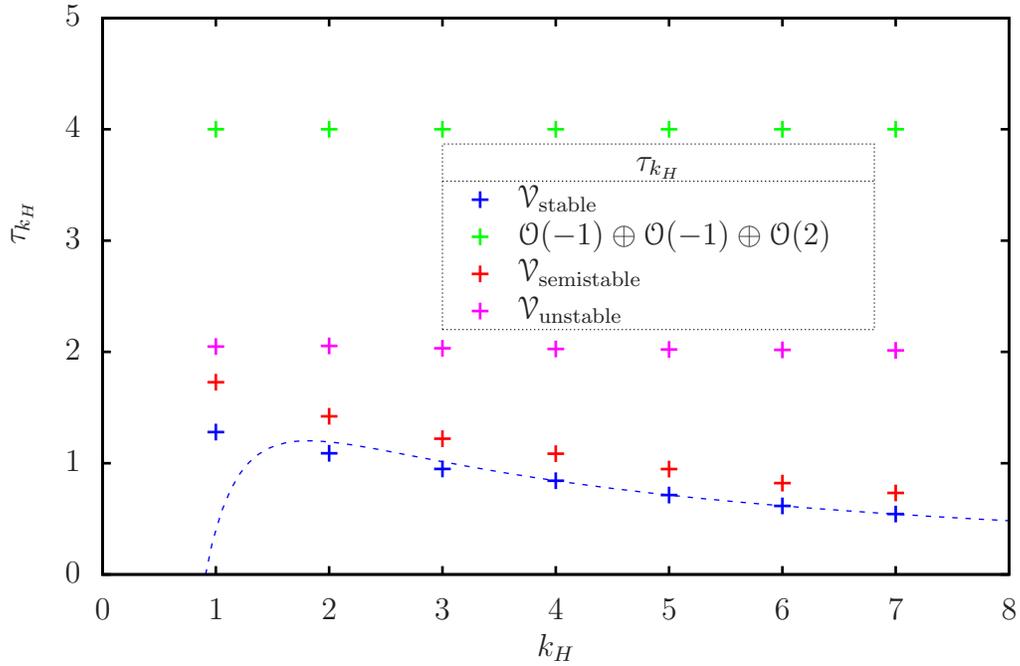}
  \end{center}
  \caption{Comparison of different bundles on the Quintic threefold.
    The stable bundle, \eqref{quintic_stable}, exhibits the behavior
    of a Hermitian Yang-Mills connection and the error $\tau_{k_H}$ decreases as the
    degree $k_H$ is increased. Meanwhile, the unstable,
    \eqref{quintic_unstable}, and strictly semi-stable,
    \eqref{quintic_semi}, bundles approach a fixed, non-zero constant
    value. The results for the semi-stable bundle depend on the choice of bundle moduli and can be made to converge to a constant arbitrarily close to zero. }
  \label{fig:CompareStableUnstableQuintic}
\end{figure}

The comparison between the stable, \eqref{quintic_stable},
semi-stable, \eqref{quintic_semi}, and the two unstable bundles, \eqref{home1} and
\eqref{quintic_unstable}, is given  in
\autoref{fig:CompareStableUnstableQuintic}.



\section{Conclusions and future work}

Donaldson's algorithm \cite{MR2161248, MR1916953, DonaldsonNumerical} has been shown to be an important  tool in the computation of Ricci-flat metrics on Calabi-Yau manifolds. Numerical implementations of this algorithm were given in \cite{Braun:2007sn,Braun:2008jp,Douglas:2006rr, Douglas:2006hz} and a generalization of these techniques to Hermitian metrics on holomorphic vector bundles was proposed in \cite{MR2154820,Douglas:2006hz}. In this paper, we have presented a systematic and efficient method to implement the generalized Donaldson algorithm and to numerically compute Hermitian Yang-Mills connections satisfying \eqref{the_hym}. 
We illustrated this by, first, computing the Ricci-flat Calabi-Yau metrics on the Quartic $K3$ surface and the Quintic threefold and, second, over these manifolds, calculating the Hermitian Yang-Mills connection on several holomorphic vector bundles defined via the monad construction \cite{okonek, Anderson:2007nc, Anderson:2008uw,Anderson:2009mh}.

In addition to showing that the algorithm converges to the Hermitian Yang-Mills connection for a slope-stable bundle, we presented results demonstrating that even in the case of unstable vector bundles, for which the generalized T-operator \eqref{t_gen} does not converge, the algorithm produces a physically relevant connection. In Section \ref{sec:StableVsUnstable}, we showed that the connection produced by the generalized Donaldson algorithm in the case of an unstable bundle
can be understood in terms of the Harder-Narasimhan filtration \cite{huybrechts} of an unstable sheaf by semi-stable subsheaves. Furthermore, as shown in Subsection \ref{geodesic}, in cases where the filtrations contain sheaves with curvature singularities, the presence of such singularities can be accurately described by the algorithm. These results may shed light on the mathematical study of non-Hermitian Yang-Mills connections \cite{kaledin-1998-4}.

Importantly, our results for unstable bundles allow us to apply the generalized Donaldson algorithm to arbitrary vector bundles arising in heterotic string compactifications and use it to determine whether such geometries admit supersymmetric vacua. The problem of deciding whether or not a given holomorphic vector bundle is slope-stable is a notoriously difficult one. Particularly challenging is the fact that the difficulty of a stability analysis generally increases rapidly with the dimension, $h^{1,1}$, of the K\"ahler cone. One of the great advantages of the algorithm presented in this paper is that, unlike in a standard analytic analysis, such numerical calculations can, in principle, be performed with essentially equal ease in arbitrary $h^{1,1}$ dimensions, thereby providing an important new tool in the analysis of supersymmetric heterotic vacua.

In recent work, \cite{Anderson:2009sw,Anderson:2009nt}, the dependence of the slope-stability property of a vector bundle on the K\"ahler and bundle moduli has been explored from the point of view of four-dimensional effective field theory. The presence of ``stability walls" separating stable and unstable regions of K\"ahler moduli space has been shown to have a variety of interesting physical and mathematical consequences \cite{Anderson:2010tc, bundle_trans}. In future work \cite{higher_cones}, we will use the numerical techniques presented here to further explore the moduli dependence of supersymmetric solutions.

Finally, it is worth noting that the results of this paper are an important step forward in the computation of observable quantities of particle physics, such as the matter-field K\"ahler potential and physical Yukawa couplings. We hope to explore these applications in the future.

\section*{Acknowledgments}

The work of L. Anderson and B. A. Ovrut is supported in part by the DOE under contract No. DE-AC02-
76-ER-03071 and by NSF RTG Grant DMS-0636606. L. Anderson, V. Braun and B. Ovrut would like to thank the Kavli Institute for Theoretical Physics for hospitality while this work was being completed.

\appendix
\makeatletter
\def\Hy@chapterstring{section}
\makeatother

\bibliographystyle{utcaps} 
\renewcommand{\refname}{Bibliography}
\addcontentsline{toc}{section}{Bibliography} 

\bibliography{Volker,Metric}

\end{document}